\documentstyle[12pt,epsf]{article}

\newcommand{\beq}{\begin{equation}}
\newcommand{\eeq}{\end{equation}}
\newcommand{\beqa}{\begin{eqnarray}}
\newcommand{\eeqa}{\end{eqnarray}}
\newcommand{\eq}[1]{(\ref{#1})}
\newcommand{\nn}{\nonumber}

\newcommand{\hs}[1]{\hspace{#1 mm}}

\newcommand{\ra}{\rightarrow}
\newcommand{\trone}{{\rm tr}[1]}
\newcommand{\sh}{{\rm sinh}}
\newcommand{\ch}{{\rm cosh}}

\newcommand{\fr}{\frac}    
\newcommand{\teta}{\Theta_2\Bigl(s{2\mu\over\beta},
is{4\pi\over\beta^2}\Bigr)}
\newcommand{\integ}{\int_0^\infty}
\newcommand{\intep}{\int_0^{\pi/2}}
\newcommand{\NP}[1]{ {\it Nucl.~Phys.} {\bf #1}}
\newcommand{\PL}[1]{ {\it Phys.~Lett.} {\bf #1}}
\newcommand{\Prep}[1]{ {\it Phys.~Rep.} {\bf #1}}
\newcommand{\PR}[1]{ {\it Phys.~Rev.} {\bf #1}}
\newcommand{\PRL}[1]{ {\it Phys.~Rev.~Lett.} {\bf #1}}

\newcommand{\MPL}[1]{ {\it Mod.~Phys.~Lett.} {\bf #1}}
\newcommand{\IJMP}[1]{ {\it Int.~J.~Mod.~Phys.} {\bf #1}}

\newcommand{\ZP}[1]{ {\it Z.~Phys.} {\bf #1}}
\newcommand{\RMP}[1]{ {\it Rev.~Mod.~Phys.} {\bf #1}}
\newcommand{\AP}[1]{ {\it Ann.~Phys.} {\bf #1}}
\begin{document}

\begin{titlepage}


\setcounter{page}{0}
\vspace{-1 mm}
\begin{center}
{\Large Thermodynamic Gross-Neveu model 
         in a constant electromagnetic field }
\end{center} 
\vspace{4 mm}
\begin{center}
{\bf  Shinya Kanemura$^{a)}$
  \footnote{e-mail: kanemu@theory.kek.jp}}, 
{\bf Haru-Tada Sato$^{b)}$
\footnote{e-mail: sato@nbi.dk}} and 
{\bf Hiroshi Tochimura$^{c)}$
\footnote{e-mail: tochi@theo.phys.sci.hiroshima-u.ac.jp}}\\
\vspace{1mm}
$^{a)}${\small \em  Theory Group, KEK,\\
       Tsukuba, Ibaraki 305, Japan,}\\
$^{b)}${\small \em The Niels Bohr Institute, University of Copenhagen, \\
      Blegdamsvej 17, DK-2100 Copenhagen, Denmark}\\
$^{c)}${\small \em Department of Physics, Hiroshima University, \\
     Kagamiyama 1-3-1, Higashi-Hiroshima 739, Japan}\\
\end{center}
\vspace{3mm}
\begin{abstract}
Analyzed is the effective potential of $D$-dimensional thermodynamic 
Gross-Neveu model (defined by large-$N$ leading order, $2\leq D<4$) 
under constant electromagnetic field ($\vec{E}\cdot\vec{B}=0$). 
The potential is derived from a thermal analogy of the worldline 
formalism. In the magnetic case, the potential is 
expressed in terms of a discretized momentum sum instead of a 
continuum momentum integral, and it reproduces known critical values 
of $\mu$ and $T$ in the continuum limit. In the electric case, 
chiral symmetry is restored at finite $T$ (with arbitrary $\mu$) 
against instability of fermion vacuum. An electromagnetic duality 
is pointed out as well. Phase diagrams are obtained in both cases. 
\end{abstract}

\vspace{2mm}
\begin{flushleft}
\end{flushleft}
\end{titlepage}

\renewcommand{\thefootnote}{\arabic{footnote}}
\section{Introduction}
\setcounter{equation}{0}
\indent

%
The Gross-Neveu (GN) model is known as the simplest model possessing 
the similar properties to QCD; dynamical breaking of (discrete) 
chiral symmetry, dimensional transmutation, asymptotic freedom and 
renormalizability ($D < 4$). Indeed, a number of GN models 
(and related models such as the NJL models) have often been studied 
as the prototypes of dynamical symmetry breaking in the various 
circumstances of finite temperature \cite{TGN,Wol,IKM}, 
electromagnetic field \cite{HPY}~-\cite{GMST} and finite 
curvature \cite{RGN,GOS}. The high temperature behaviors of these 
models \cite{Kli3,GMST,NJLT,KS} may potentially be interesting for 
exploring a dynamical symmetry breaking in (quark-gluon) plasma and 
in cosmological situations like early universe or a neutron star. 

In this paper, we examine the effective potential and chiral phase 
structures of the $D$-dimensional ($2 \leq D <4$) GN model defined 
by the leading order of $1/N$-expansion in a magnetic $B$ 
(or electric $E$) background field at finite temperature and density. 
In the case of $D$ being slightly below 4, the plasma physics is an 
interesting application \cite{NJLT}. 
In the case of $D=3$ \cite{GMST}, the physics of our 
model concerns planar condensed matter systems such as BCS theory: 
a magnetic field works as a catalyst for generating an energy gap in 
fermion spectra, through enhancing the interactions of fermions at 
small momenta even if the weakest (attractive) interaction. Gusynin 
et. al. suggested \cite{GMST,QED} that this effect should be universal, 
i.e., model independent, in $D=3$ and should be considered in the 
phenomena such as quantum Hall effects and high temperature 
superconductivity. Similarly in a supersymmetric model, constant 
magnetic field generates fermion's dynamical mass \cite{SNJL}.  

In the case of $D=2$, magnetic field can not exist, 
however the GN model by itself corresponds to a model of 
poly-acetylene. The first order phase transition of poly-acetylene 
from solitonic (broken) phase to metallic (symmetric) phase is 
formulated~\cite{CM} as a function of doping concentration (chemical 
potential) within the large-$N$ approximation. The critical value of 
chemical potential coincides with an experimental value, and 
recently a $1/N$-correction to this result was obtained in 
\cite{CM2}. Our study might explain some thermodynamic behaviour 
of a poly-acetylene model under constant electric field in a 
finite temperature and density system. 

Our crucial interest is therefore the question how universal the 
effect of magnetic catalyst on phase structures is in the setting 
of both finite temperature $T$ and chemical potential $\mu$ in 
various dimensions. We study this question for an electric background 
field as well. In order to deal with these models all together, it is, 
of course, useful to analyze a dimensionally regulated effective 
potential. Such approaches were done in the cases of finite 
temperature \cite{IKM}, curvature \cite{RGN} and their 
mixture \cite{KS}.  We shall derive a universal expression of 
the effective potential as a function of $D$, $T$, $\mu$ and the 
external field parameter $\xi$ ($=e\sqrt{|\vec{B}|^2-|\vec{E}|^2}$, 
where we have assumed $\vec{B}\cdot\vec{E}=0$) within the leading 
order of the $1/N$-expansion. We shall then present various 
interesting properties which can be derived from the universal 
effective potential. Although we have not completed our analyses yet, 
we confirm that our present results will be useful in order to 
establish a foundation for further developments in the various 
branches mentioned above. 

This paper is organized as follows. 
In Sect.2, we present a simple and convenient derivation of the 
effective potential. There is no manifestation of exact fermion 
propagators of finite temperature field theory in the derivation. 
Namely, we do not use its Feynman rules to evaluate a fermion 
loop. Indeed, we apply an analogy of the worldline formalism 
\cite{Str},\cite{SS} to the finite temperature cases. Then we 
write down the proper-time integral representations for the 
critical equations which determine the phase structures in 
the $\xi$-$T$-$\mu$ space. 

In Sect.3, we discuss the cases of magnetic field dominant, 
i.e., $\xi$ is a real number. In the former half of this section, 
we analyze the zero-temperature limit of the effective potentials 
for finite $\xi$ and $\mu$. The potential is shown to be a 
discretized version of the one presented in \cite{IKM}. 
Also, first order phase transitions are observed in large $\mu$ 
regions on the $\xi$-$\mu$ ($T=0$) plane. 
In the latter half of the section, we discuss the phase structures. 
We find that a constant magnetic field works as the catalyst in 
$3 \leq D <4$ at finite $T$ and $\mu$, and verify that the magnitude 
of catalyst effect becomes smaller as $D$ or $\mu$ increasing. 

In Sect.4, we present the electric dominant phase diagrams, i.e., 
the cases of $\xi$ a pure imaginary number. Analyzing the imaginary 
part of the effective potential, we show that chiral symmetry can be 
restored in high $T$ situations ($\mu\geq0$) in $2\leq D<4$, with 
vanishing imaginary part. We also show that some analytic results 
at $T=\mu=0$ can be obtained through the electromagnetic dual 
rotation $\xi\ra i\rho$. 

Conclusions and discussions are in Sect.5. 
In Appendices, some details on the $\xi\ra0$ limit are 
presented. One can find how to reproduce various 
known results at $\xi=0$ \cite{Wol,CM,IKM} from our results 
in this continuum limit $\xi\ra0$.

\section{The effective potential in $\xi$-$T$-$\mu$ space}
\setcounter{equation}{0}
\indent


The model Lagrangian is 
\begin{eqnarray}
  {\cal L} = \bar{\psi} i \gamma^{\mu} D_{\mu} \psi 
           - \fr{N}{2 \lambda} \sigma^2 - \sigma \bar{\psi} \psi, 
\label{a}
\end{eqnarray}
where $\sigma$ is the auxiliary field and $N$ the number of flavors. 
The fermions $\psi = \{ \psi_i \}, (i = 1, \cdot\cdot\cdot, N)$, are 
coupled to the background gauge fields via the covariant derivative 
$D_{\mu} = \partial_\mu - i e A_{\mu}$. The effective 
potential is given by 
\beq
V(\sigma,\xi) = V(\sigma)-V(0)\ ,      \label{poten2}
\eeq                                       
with
\beq
V(\sigma)={1\over2\lambda}\sigma^2 - 
       \ln {\rm Det}(i\gamma^\mu D_\mu-\sigma).   \label{lndet}
\eeq
This model does not possess a continuum chiral symmetry but a 
discrete one. After the effective potential is obtained as a 
function of the auxiliary field, the potential becomes the same 
form as that of the NJL model.

We shall show below that the $D$-dimensional 
thermodynamic effective potential is given by 
\begin{eqnarray}
  V(\sigma;\xi,\beta,\mu)=
     {1\over2\lambda}\sigma^2+{\trone\over2\beta}
       \integ {ds\over s}{\teta\over(4\pi s)^{(D-1)/2}}
     s\xi {\rm coth}(s \xi) e^{s\mu^2} (e^{-s\sigma^2}-1), \label{b} 
\end{eqnarray}  
where $\Theta_2$ is the elliptic theta function of second kind and 
$\trone$ means the trace of gamma matrix unit (we do not need 
its explicit value). The $D=3$ (magnetic dominant) case was derived 
in \cite{GMST}. Also, by letting $T (=\beta^{-1})$ and $\mu$ to be 
zero, this potential reproduces the potentials derived in the $D=3$ 
magnetic dominant (real $\xi$) cases \cite{Kli1,Kli3} and the electric 
dominant (pure imaginary $\xi$) cases in $D=2$ \cite{KKM,Stone}, 
$D=3$ \cite{Kli2} and $D=4$ \cite{KLL}. 

\subsection{Derivation of the effective potential}
\indent

In this subsection, we present a somewhat different derivation of 
the effective potential from that done in \cite{GMST}, where they 
evaluated it in the Matsubara (imaginary time) method using a 
fermion propagator under a constant magnetic field. We shall derive 
the  potential \eq{b}, modifying the worldline formalism \cite{Str,SS} 
into the thermodynamic case. 

Let us start with reviewing a few notes on $V(\sigma)$ for 
convenience of notation. 
In the Schwinger proper time method \cite{PT}, a basic expression 
for the potential \eq{lndet} already exists in \cite{HPY} 
\beq
  V(\sigma)=
     {1\over2\lambda}\sigma^2+{1\over2}{\rm tr} \integ {ds\over s}
{1\over(4\pi s)^{D/2}}
 e^{-s\sigma^2} \exp\Bigl[-{1\over2}esF_{\mu\nu}\sigma^{\mu\nu} 
-{1\over2}{\rm Tr}\ln \left({\sin(esF)\over esF}\right)_{\mu\nu}\Bigr],
\label{potential}
\eeq             
where tr and Tr mean the traces in the Gamma and Lorentz 
matrix spaces respectively. The trace of $F^{2n}_{\mu\nu}$ is given 
by the following in respective dimensions: $2E^{2n}$ in $D=2$, 
$2(\vec{E}^2-B^2)^n$ in $D=3$ and $2[(\rho_+^2)^n + (\rho_-^2)^n]$ 
in $D=4$, where 
\[
\rho_\pm^2 = {1\over2}\Bigl[\,\vec{E}^2-\vec{B}^2\pm 
\sqrt{(\vec{B}^2-\vec{E}^2)^2 +4(\vec{E}\cdot\vec{B})^2}\,\,\Bigr].
\]
If we assume $\vec{E}\cdot\vec{B}=0$ in $D=4$, and if the notation 
$\rho=e\sqrt{\vec{E}^2-\vec{B}^2}$ is properly understood in each 
case (for example, $\vec{B}$ should be zero in $D=2$), we can 
write those trace values in a unified form 
\beq
{\rm Tr}\Bigl[(eF)^{2n}\Bigr]_{\mu\nu} = 2 \rho^{2n}. \label{treF}
\eeq
Similarly, the eigenvalue of ${1\over2}e\sigma\cdot F$ can be 
expressed as $\pm i\rho$ multiplied by the degeneracy factor 
${1\over2}\trone$ in each $D$, where we have again used 
$\vec{E}\cdot\vec{B}=0$ in $D=4$. In this way, we re-write
\beqa 
& &{\rm tr}\exp\Bigl[-{1\over2}esF_{\mu\nu}\sigma^{\mu\nu}\Bigr] 
            = \trone\ch(s\xi),  \\
& &{\rm Tr}\Bigl[\ln{\sin(esF)\over esF} \Bigr]_{\mu\nu}
            = 2\ln{\sh(s\xi)\over s\xi},           \label{trsine}
\eeqa
where $\xi=i\rho=e\sqrt{\vec{B}^2-\vec{E}^2}$. Note that $\xi$ can not 
be real in $D=2$ because $\vec{B}=0$. Thereby the potential 
\eq{potential} becomes \cite{KKM,KLL,Kli1,Kli2,GMST} 
\beq
  V(\sigma)=
     {1\over2\lambda}\sigma^2+{1\over2}\trone \integ {ds\over s}
{1\over(4\pi s)^{D/2}}
 e^{-s\sigma^2} s\xi \coth(s\xi). 
\label{potential2}
\eeq 
Note that the external field dependence 
$s\xi\coth s\xi$, the loop momentum contribution $(4\pi s)^{-D/2}$ 
and the mass term $e^{-s\sigma^2}$ are decoupled in this expression. 
 
Our proposal for introducing temperature and chemical potential is 
the replacement of the (normalization) factor $(4\pi s)^{-D/2}$ 
with a thermodynamic normalization. To see this, 
let us consider an alternative derivation of the second term on 
RHS of \eq{potential2}. It is the same as the one-loop effective 
action of a spinor loop (with mass $\sigma$) coupled to a background 
gauge (constant field strength) $A_\mu={1\over2}x^\nu F_{\nu\mu}$. 
In the worldline formalism, the effective action is organized in a 
form of closed path integral of one-dimensional Euclidean 
action \cite{Str,SS}
\beqa
\Gamma[A]&=&\ln {\rm Det}(i\gamma^\mu D_\mu -\sigma) \nn \\
&=& {1\over2}\integ{ds\over s}e^{-\sigma^2s}\oint{\cal D}x{\cal D}
\psi \nn\\
&&\times\exp\Bigl[-\int_0^s d\tau({1\over4}{\dot x}^2
+ {1\over2}\psi{\dot\psi} + {1\over2}iexF{\dot x} 
- ie\psi F\psi)\,\Bigr],
\eeqa
where $x=x^\mu(\tau)$ and $\psi=\psi^\mu(\tau)$ are the periodic and 
anti-periodic world-line (real) fields which are defined on a closed 
loop parametrized by $\tau$ ($0\leq\tau\leq s$), and ${\dot x}$ means 
$\partial_\tau x(\tau)$ etc. The closed path integral amounts to 
the super determinant
\beq
{\rm SDet}^{-1/2}\left( 
\begin{array}{ll}
 -{1\over4}\partial^2_\tau +{1\over2}ieF\partial_\tau
                               &\quad\quad 0  \\
 0 &\quad\quad  {1\over2}\partial_\tau-ieF
\end{array}\right)
={\cal N}\left[ 
          { {\rm Det}(1-2ieF\partial^{-1}_\tau)_P \over
            {\rm Det}(1-2ieF\partial^{-1}_\tau)_A } \right]^{-1/2},
\eeq
where ${\cal N}$ is defined by the free part integral \cite{Str}, 
\beq
{\cal N}=\oint{\cal D}x{\cal D}\psi 
\exp\Bigl[-\int_0^s({1\over4}{\dot x}^2 
+{1\over2}\psi{\dot\psi})d\tau\,\Bigr],
\eeq
and its explicit value is given by the formulae
\beq
\oint{\cal D}x\exp\Bigl[-\int_0^s{1\over4}{\dot x}^2d\tau\,\Bigr]
=\int{d^Dk\over(2\pi)^D}e^{-sk^2} =(4\pi s)^{-D/2},  \label{norm}
\eeq
\beq
\oint{\cal D}\psi\exp\Bigl[-\int_0^s{1\over2}\psi{\dot \psi} 
d\tau\,\Bigr]=-\trone\ . \label{normf}
\eeq
Thus the effective action yields 
\beq
\Gamma[A]=
  -{1\over2}\trone\int{ds\over s}(4\pi s)^{-D/2}e^{-\sigma^2s}
  \left[ { {\rm Det}(1-2ieF\partial^{-1}_\tau)_A \over
            {\rm Det}(1-2ieF\partial^{-1}_\tau)_P } \right]^{1/2}\ . 
\eeq
The determinants can be evaluated on the Fourier bases 
$\{\exp[2\pi ik\tau/s]|k\in{\bf Z}, k\not=0\}$ for the periodic 
and $\{\exp[2\pi i(k+{1\over2})\tau/s]|k\in{\bf Z}\}$ for the 
anti-periodic functions \cite{SS}. Note that the zero mode in the 
periodic basis is excluded. We have 
\beq
{\rm Det}\Bigl(1-2ieF\partial^{-1}_\tau\Bigr)_P=
{\rm Det}'\prod_{k=-\infty,\not=0}^{\infty}
\Bigl(1-2ieF{s\over2\pi ik}\,\Bigr)={\rm Det}'{\sin(eFs)\over eFs},       
\eeq
and similarly
\beq
{\rm Det}\Bigl(1-2ieF\partial^{-1}_\tau\Bigr)_A = 
{\rm Det}'\prod_{k=-\infty}^{\infty}
\Bigl(1-2ieF{s\over2\pi i(k+1/2)}\,\Bigr)=
{\rm Det}'\cos(eFs), 
\eeq
where ${\rm Det}'$ concerns the determinant on a Lorentz matrix. 
We then arrive at the expression \cite{SS}
\beq
\Gamma[A]=
   -{1\over2}\trone\int{ds\over s}(4\pi s)^{-D/2}e^{-\sigma^2s}
  {{\rm Det}'}^{1\over2}\Bigl[eFs\cot(eFs)\Bigr],
\eeq      
and this coincides with the second term of \eq{potential2} with 
the identification of \eq{treF}. 

Now let us proceed to the thermodynamic case. We have to replace 
$k_0$-integral according to the Matsubara formalism
\beq
k_0 \ra {\pi\over\beta}(2n+1) -i\mu, \qquad\quad
\int{dk_0\over2\pi} \ra {1\over\beta}\sum_n.
\eeq
Namely, we change the path-integral normalization for $x_0$ 
through discretizing $k_0$-integral in \eq{norm}
\beqa
\oint{\cal D}x_0\exp\Bigl[-\int_0^s{1\over4}{\dot x}_0^2\,\Bigr]
&\ra& {1\over\beta}\sum_{n=-\infty}^{\infty}
\exp\Bigl[-s\{{\pi\over\beta}(2n+1)-i\mu\}^2\,\Bigr] \nn\\
&=&{1\over\beta}e^{s\mu^2} 
\Theta_2(s{2\mu\over\beta},is{4\pi\over\beta^2}),  \label{norm2}
\eeqa
and \eq{norm} is then replaced by 
\beq
\oint_{\beta,\mu}{\cal D}x\exp\Bigl[-\int_0^s{1\over4}{\dot x}^2\,\Bigr] 
=(4\pi s)^{-{D-1\over2}}{1\over\beta}e^{s\mu^2} 
\Theta_2(s{2\mu\over\beta},is{4\pi\over\beta^2}). \label{normbm}           
\eeq
Since the zero temperature limit $\beta\ra\infty$ of 
RHS of \eq{norm2} is 
\beq
\lim_{\beta\ra\infty}{1\over\beta}e^{s\mu^2} 
\Theta_2(s{2\mu\over\beta},is{4\pi\over\beta^2})=
{1\over\sqrt{4\pi s} }\ ,         \label{zeroT}
\eeq
\eq{norm} is reproduced from \eq{normbm}. Equivalently, this can be 
verified by decoupling the zero temperature normalization part 
$(4\pi s)^{-D/2}$,
\beqa
\oint_{\beta,\mu}{\cal D}x\exp\Bigl[-\int_0^s{1\over4}{\dot x}^2\Bigr]  
&=& \oint{\cal D}x\exp\Bigl[-\int_0^s{1\over4}{\dot x}^2\Bigr] \nn\\
&+& 2(4\pi s)^{-D\over2}\sum_{n=1}^\infty (-1)^n 
e^{-n^2\beta^2/4s}\cosh(n\beta\mu)\ , \label{betu}
\eeqa
where we have applied the transformation 
\beqa
e^{s\mu^2}{\sqrt{4\pi s}\over\beta}\teta &=& 
\Theta_4(i{\beta\mu\over2\pi},i{\beta^2\over4\pi s}) \nn\\
&=& 1+2\sum_{n=1}^\infty (-1)^n e^{-n^2\beta^2/4s}\cosh(n\beta\mu).
\label{transj}
\eeqa 
Obviously, the second term in \eq{betu} vanishes in $\beta\ra\infty$, 
and is \eq{norm} recovered.                                

Let us remember that the thermal parameter $\beta^{-1}$ has nothing 
to do with the loop length $s$ which is the Schwinger proper time of 
the spinor loop \cite{PT}. All integrations on internal momenta 
$(k_i,k_0)$ are now done, and thus the operator determinants do not 
change any more, since $\partial_\tau$ acts on the proper time 
coordinate and $F$ is a constant matrix. We therefore conclude 
\beq
\Gamma[A]=
   -{1\over2}\trone\int{ds\over s}
{ \Theta_2(s{2\mu\over\beta},is{4\pi\over\beta^2}) 
\over (4\pi s)^{(D-1)/2} \beta}
e^{s(\mu^2-\sigma^2)} s\xi\coth(s\xi).
\eeq      
Replacing the logarithmic part of \eq{lndet} by this quantity, 
we get 
\beq
  V_{\beta,\mu}(\sigma;\xi)=
   {1\over2\lambda}\sigma^2+{\trone\over2\beta}
   \integ {ds\over s}{\teta\over(4\pi s)^{(D-1)/2}}
  s\xi {\rm coth}(s \xi) e^{-s(\sigma^2-\mu^2)}.  \label{potential3}
\eeq  
When $D=3$, this thermodynamic potential coincides with the 
equation (B6) of Ref.\cite{GMST}. Also, when $\xi=0$, 
$V_{\beta,\mu}(\sigma;\xi=0)$ coincides with the $D$-dimensional 
thermodynamic potential discussed in \cite{IKM}. When $T=\mu=0$, 
$V_{\beta,\mu}(\sigma;\xi)$ is reduced to \eq{potential2} because 
of \eq{zeroT}. Finally, subtracting the zero point value of 
the potential, we obtain the announced result \eq{b}
\beq
  V(\sigma;\xi,\beta,\mu) = 
V_{\beta,\mu}(\sigma;\xi) - V_{\beta,\mu}(0;\xi).   \label{poten1}
\eeq
\subsection{The critical equations}
\indent

%
In the following, we list the critical equations that are entire 
bases of our numerical analyses of chiral phase structures. 
We should remember that the original effective potential \eq{lndet} 
contains the ultra-violet divergence and also has an imaginary part 
for pure imaginary $\xi$. When the imaginary part disappears, we 
shall impose the following renormalization condition 
(on the real part) 
\begin{eqnarray}
\left. 
\lim_{T,\mu \rightarrow 0} {\partial^2 \over \partial \sigma^2}
{\rm Re}\,V(\sigma;\xi,T,\mu) 
\right|_{\sigma=1} = \fr{1}{\lambda_R},  
\end{eqnarray}
and we can thus define the renormalized coupling constant 
$\lambda_R$ in the following way 
\begin{eqnarray}
\fr{1}{\lambda}- \fr{1}{\lambda_R} = \trone{\rm Re}\,
\int_0^\infty \fr{ds}{(4\pi s)^{D/2}} e^{-s}(1-2s)\, 
s \xi {\coth}(s\xi). \label{d} 
\end{eqnarray}
The real part of renormalized effective 
potential is then expressed in terms of $\lambda_R$ as 
\begin{eqnarray}
{\rm Re}\,V_R(\sigma;\xi,T,\mu) &=& \fr{1}{2\lambda_R} \sigma^2 
+ \fr{1}{2}\trone{\rm Re}\,\int_0^{\infty} \fr{ds}{(4\pi s)^{D/2}} 
s \xi {\coth}(s\xi)    \label{repotential}  \\ 
&\times& \Bigl[ \fr{1}{s} (e^{-s\sigma^2}-1) \fr{\sqrt{4\pi s}}{\beta}
e^{s\mu^2}\teta + \sigma^2 e^{-s}(1-2s) \Bigr]\ .  \nn
\end{eqnarray}
The gap equation to determine the dynamical mass, i.e., the lowest 
energy configuration of $\sigma$, is then 
\begin{eqnarray}
0&=& \fr{1}{\lambda_R}+\trone{\rm Re}\,\int_0^{\infty}
   \fr{ds}{(4\pi s)^{D/2}} s \xi {\coth}(s\xi)   \nn\\
 &&\ \times\Bigl[ -e^{-s(\sigma^2-\mu^2)}
     \fr{\sqrt{4\pi s}}{\beta}\teta +e^{-s}(1-2s)\, \Bigr]\ ,  
 \label{gap}
\end{eqnarray}
where we have eliminated the trivial solution $\sigma=0$. 

Now let us derive the equations to determine phase diagrams 
in the $\xi$-$T$-$\mu$ parameter space. The critical surface of 
first order is governed by the simultaneous equations 
\beq
{\rm Re}\, V_R (\sigma=m, \xi, \beta, \mu) = 0, \label{cri1order1}
\eeq
\beq
\left.\fr{\partial}{\partial\sigma}{\rm Re}\,
V_R(\sigma,\xi,\beta,\mu) \right|_{\sigma =m} = 0,
   \label{cri1order2}
\eeq
where $m$ denotes a non-trivial solution of \eq{gap}. Combining these 
equations, we obtain 
\beqa
0&=&{\rm Re}\,\integ{ds\over(4\pi s)^{D/2}}s\xi\coth(s\xi)\cdot
{\sqrt{4\pi s}\over\beta}e^{s\mu^2}\teta \nn\\
&&\ \times\Bigl[\,m^2e^{-sm^2} +{1\over s}(e^{-sm^2}-1)\,\Bigr]\ .
\label{c1equation}
\eeqa
One can decouple RHS of this equation into $V(m;\xi)$ and a pure 
thermal part using \eq{transj}, however we stop writing further 
details, which will be discussed in sect.3.1.
The necessary condition to determine the critical surface of 
second order is generally expressed by 
\begin{eqnarray}
  \lim_{\sigma \rightarrow 0} \fr{\partial}{\partial \sigma^2}
{\rm Re}\,V_R(\sigma, \xi, \beta, \mu) = 0\ ,        \label{cond1}
\end{eqnarray}  
and this reads 
\beqa
0&=&{\rm Re}\,\int_0^{\infty}\fr{ds}{(4\pi s)^{D/2}} 
\Bigl[ s \xi {\coth}(s\xi)
  \Bigl\{ -e^{s\mu^2}\fr{\sqrt{4\pi s}}{\beta}\teta \nn \\
&&\hskip90pt +\,e^{-s}(1-2s) \,\Bigr\}+ 2se^{-s} \Bigr]\ . \label{j} 
\eeqa
Here, we have adopted the following value of the renormalized 
coupling constant
\begin{eqnarray}
\fr{1}{\lambda_R}=
      \trone\int_0^{\infty} 
      \fr{ds}{(4\pi s)^{D/2}}2se^{-s},    \label{coupling}
\end{eqnarray}
which makes the chiral symmetry broken at $T=\mu=\xi=0$ with 
generating the dynamical mass $\sigma=1$. Note that a solution 
of \eq{j} does not always represent a true second order transition 
point if a first order transition occurs. We have to cut an 
irrelevant piece away from the true critical surface solving the 
equation of third order critical line. The third order critical 
line, which yields the boundary between the first and second order 
critical surfaces, is determined by the simultaneous equations 
\eq{cond1} and 
\begin{eqnarray}
  \lim_{\sigma \rightarrow 0} 
\left( \fr{\partial}{\partial \sigma^2} \right)^2 
{\rm Re}\,V_R(\sigma, \xi, \beta, \mu) = 0\ .        \label{cond2}
\end{eqnarray}
The latter equation leads to 
\begin{eqnarray} 0 =
 {\rm Re}\, \int_0^\infty \fr{sds}{(4\pi s)^{D/2}}
  s \xi {\rm coth} (s \xi) e^{s\mu^2} \fr{\sqrt{4\pi s}}{\beta}
  \Theta_2 \left(s \fr{2\mu}{\beta}, i s \fr{4\pi}{\beta^2} \right).
\label{m}
\end{eqnarray}

One can perform any combination of limits $\xi, T$ and/or $\mu\ra0$ 
in the 2'nd and 3'rd order critical equations \eq{j} and \eq{m}. 
{}~For instance, taking the limit $\xi \rightarrow 0$ in each 
equation, we obtain the following analytic equations 
respectively \cite{KS,IKM} (see Appendix A) 
\begin{eqnarray}
\beta^{D-2}\Gamma (1-\fr{D}{2})=\fr{2}{\sqrt{\pi}}(2\pi)^{D-2} 
\Gamma (\fr{3 - D}{2}) {\rm Re}\zeta 
(3 - D, \fr{1}{2} + i \fr{\beta \mu}{2\pi}),   \label{c2line}
\end{eqnarray}
and 
\begin{eqnarray}
  {\rm Re} \zeta (5 -D, \fr{1}{2} + i \fr{\beta \mu}{2\pi}) = 0, 
\label{tricripoint}
\end{eqnarray}
where $\zeta$ is the generalized zeta function. It is shown in 
\cite{IKM} that these equations reproduce the critical equations 
derived in \cite{Wol}. The limit $\beta\ra\infty$ can also be taken 
by applying \eq{zeroT} to \eq{j} and \eq{m}, and their limits are 
valid unless any 1'st order phase transition takes place at $T=0$.  
In contrast with these cases, the effective potential on the 
$\xi$-$\mu$ plane (and hence the 1'st order critical equation there) 
can not be obtained for $\mu\not=0$ by naively applying \eq{zeroT}. 
In order to reproduce a precise limit, we have to collect all 
contributions from under the Fermi surface at $T=0$. This can be 
achieved by an insertion of the step function $\theta(\mu-k)$ in a 
loop momentum integration \cite{CM} at a very beginning of 
computation. This prescription is, however, unclear in our present 
formulation, since we already integrated the loop momenta. 
We observe a method how to deal with this problem 
in the next section. 

\section{The magnetic dominant cases}
\setcounter{equation}{0}
\indent 

%
In this section, we discuss chiral phase structures in the magnetic 
dominant cases for finite $T$ and $\mu$. We formally include $D=2$ 
case also, because it will be useful when $\xi$ is rotated to 
$i\rho$ and because it is convenient to see a $D$-dependence of 
magnetic catalyst in the region $2<D<4$. 
Since $\xi$ is real, the effective 
potential does not possess an imaginary part, i.e.,
\beq
  {\rm Re}\, V_R(\sigma;\xi,\beta,\mu) = 
             V_R(\sigma;\xi,\beta,\mu), 
\eeq
and there is no more intricacy of critical equation analysis 
(up to the $T\ra0$ limit). 
In \cite{GMST}, the $D =3$ case ($\mu=0$, $T\not=0$) 
with a constant magnetic field was analyzed, and the magnetic field 
turns out to play a role of strong catalyst of chiral 
symmetry breaking. We shall observe this feature in the 
$\xi$-$T$-$\mu$ space for various values of $D$. 
Each critical surface is determined by numerically analyzing 
the 1'st, 2'nd and 3'rd order critical equations presented 
in the previous section. As to the zero temperature limit, 
expected from \cite{Naf}, we shall show that first order phase 
transitions exist on the $\xi$-$\mu$ plane at large values 
of $\mu$. 

\subsection{The effective potential on $\xi$-$\mu$ plane} 
\indent

First we study the zero temperature limit for finite $\xi$ and $\mu$. 
As mentioned at the end of sect.2, the special care is necessary 
to analyze the effective potential (not gap equation) on the 
$\xi$-$\mu$ plane. In addition, we have to remember that $\xi^{-1}$ 
is relevant to the magnetic length $eB$ (or $\xi$ the discrete 
momentum unit), which is known as the radius of cyclotron motion of 
an electron. This assures the magnetic lattice structure, although 
it can not be seen in the theta function representations of the 
effective potential \eq{repotential} etc. 

The effective potential \eq{poten1} with the renormalization 
\eq{d} can be written as the sum of the zero-temperature potential 
\eq{poten2} and the pure thermodynamic remainder ${\tilde V}$ by 
means of the formula \eq{transj}:
\beq
V(\sigma;\xi,\beta,\mu) = V(\sigma;\xi) 
+{\tilde V}(\sigma;\xi,\beta,\mu)\ ,
\eeq
or (q.v. \eq{poten2} and \eq{poten1})
\beq
V_{\beta,\mu}(\sigma;\xi) = 
V(\sigma) +{\tilde V}_{\beta,\mu}(\sigma;\xi),
\eeq
where
\beq
V(\sigma)={1\over2\lambda_R}\sigma^2 + {\trone\over2}\integ 
{ds\over(4\pi s)^{D/2}}s\xi\coth(s\xi)\Bigl[\,{1\over s}
e^{-s\sigma^2} +\sigma^2e^{-s}(1-2s)\,\Bigr]\ ,     \label{vs}
\eeq
\beq
{\tilde V}_{\beta,\mu}(\sigma;\xi)=\trone\integ
{ds\over(4\pi s)^{D/2}}\xi\coth(s\xi)\cdot e^{-s\sigma^2}
\sum_{n=1}^\infty(-1)^n e^{-n^2\beta^2/4s}\cosh(n\beta\mu)\ .
\label{vsb}
\eeq
Using the following formulae
\beq
\coth(s\xi)=1+2\sum_{l=1}^\infty e^{-2ls\xi},  \label{coth}
\eeq
\beq
\integ ds s^{\nu-1}\exp\Bigl[-{\beta\over s}-\gamma s\Bigr]
=2 \left({\beta\over\gamma}\right)^{\nu\over2}
K_\nu\Bigl(2\sqrt{\beta\gamma}\,\Bigr),
\eeq
we can perform the $s$-integrations in \eq{vs} and \eq{vsb} just 
similarly to the $D=3$ case \cite{GMST}. The results are 
\beqa
V(\sigma)={\sigma^2\over2\lambda_R}&+&{\trone\xi\over2(4\pi)^{D/2}}
\left[2(2\xi)^{{D\over2}-1}\zeta(1-{D\over2},1+{\sigma^2\over2\xi})
\Gamma(1-{D\over2}) + \sigma^{D-2}\Gamma(1-{D\over2})\right. \nn\\          
& &\ +2\sigma^2(2\xi)^{{D\over2}-2}\zeta(2-{D\over2},1+{1\over2\xi})
\Gamma(2-{D\over2}) + \sigma^2\Gamma(2-{D\over2}) \nn\\          
& &\ \left.-4\sigma^2(2\xi)^{{D\over2}-3}
\zeta(3-{D\over2},1+{1\over2\xi})\Gamma(3-{D\over2}) 
- 2\sigma^2\Gamma(3-{D\over2}) \,\right]
\label{magpote}
\eeqa
and   
\beq
{\tilde V}_{\beta,\mu}(\sigma;\xi)={\trone\xi\over(4\pi)^{D/2}}
\,\Bigl[\,{1\over2}{\cal O}_\beta(\sigma)  
+ \sum_{l=1}^\infty{\cal O}_\beta(\sqrt{\sigma^2+2l\xi}\,)\,\Bigr]\ ,
\label{discpote}
\eeq
where we have introduced the compact notation 
\beqa
{\cal O}_\beta(\sigma) &\equiv& 2\integ ds s^{-D/2}
\sum_{n=1}^\infty(-1)^n\cosh(n\beta\mu)
\exp\Bigl[ -(s\sigma^2+{n^2\beta^2\over 4s}) \Bigr] \nn \\
&=& 4\sum_{n=1}^\infty(-1)^n\cosh(n\beta\mu)
\left({\beta n\over2\sigma}\right)^{1-D/2}K_{D/2-1}(n\beta\sigma)\ . 
\label{obeta}
\eeqa
Although we have performed the integrations w.r.t. the proper time $s$, 
it is rather convenient to consider an integral representation in order 
to observe the low temperature limit of ${\cal O}_\beta(\sigma)$. Using 
the formula 
\beq
K_\nu(z)=
{\sqrt{\pi}\left({z\over2}\right)^\nu\over\Gamma(\nu+{1\over2})} 
\int_1^\infty e^{-zt}(t^2-1)^{\nu-{1\over2}} dt\ , \quad 
\mbox{Re}\,\nu>-{1\over2}\ ,\,\,\,\mbox{Re}\,z>0\ ,
\eeq
we get the following integral representation
\beq
{\cal O}_\beta(\sigma)={4\sqrt{\pi}\over\Gamma({D-1\over2})}
\sigma^{D-2}\sum_{n=1}^\infty (-)^n \ch(n\beta\mu)\int_1^\infty 
e^{-n\beta\sigma t}(t^2-1)^{D-3\over2} dt \ .
\eeq
Performing the summation over $n$ and changing the integration 
variable $t=1+t'/\sigma$, we obtain
\beq
{\cal O}_\beta(\sigma)={-2\sqrt{\pi}\over\Gamma({D-1\over2})}\,
\Bigl[\,\int_0^\infty {(t^2+2\sigma t)^{D-3\over2}\over
1+e^{\beta(t+\sigma+\mu)}}dt\quad +(\mu\ra\mu)\,\Bigr]\ .
\eeq
${\cal O}_\beta(0)$ is given by (a derivation is in Appendix A)
\beq
{\cal O}_\beta(0) = {2\over\sqrt{\pi}}\left({2\pi\over\beta}\right)
^{D-2}\Gamma\left({3-D\over2}\right)\mbox{Re}\,\zeta\Bigl(3-D,
{1\over2}+i{\beta\mu\over2\pi}\Bigr)\ .         \label{obeta0}
\eeq

Here, we stress that \eq{discpote} is a discrete version of the 
$\xi=0$ potential obtained in \cite{IKM} 
\beq
{\tilde V}(\sigma;0,\beta,\mu)=-{\trone2\sqrt{\pi} \over
(4\pi)^{D/2}\Gamma({D-1\over2})\beta} \integ dk k^{D-2} 
\ln{ (1+e^{-\beta(\mu+\sqrt{\sigma^2+k^2})}) 
(1+e^{-\beta(\sqrt{\sigma^2+k^2}-\mu)})
\over (1+e^{-\beta(\mu+k)}) (1+e^{-\beta(k-\mu)}) } \ .
\label{contpote}
\eeq                              
This can be easily understood in the case of $D=3$, since 
${\cal O}_\beta(\sigma)$ takes the following simple form 
\beq
{\cal O}_\beta(\sigma) = -2{\sqrt{\pi}\over\beta} \ln
\Bigl[\,(1+e^{-\beta(\sigma+\mu)})
(1+e^{-\beta(\sigma-\mu)})\,\Bigr] \ ,
\eeq
and hence we have (see also \cite{GMST,Anders})
\beqa
{\tilde V}(\sigma;\xi,\beta,\mu)
&=& {\tilde V}_{\beta,\mu}(\sigma;\xi) - 
    {\tilde V}_{\beta,\mu}(0;\xi) \nn\\
&=& -{\trone\xi2\sqrt{\pi}\over(4\pi)^{3/2}\beta}\Bigl[
{1\over2}\ln{ (1+e^{\beta(\mu-\sigma)}) (1+e^{-\beta(\mu+\sigma)})
\over (1+e^{\beta\mu}) (1+e^{-\beta\mu}) } \nn\\
&+& \sum_{l=1}^\infty
\ln{ (1+e^{\beta(\mu-\sqrt{\sigma^2+2l\xi})}) 
(1+e^{-\beta(\mu+\sqrt{\sigma^2+2l\xi})})
\over (1+e^{\beta(\mu-\sqrt{2l\xi})}) 
      (1+e^{-\beta(\mu+\sqrt{2l\xi})}) } \Bigr].  \label{dpote}
\eeqa
This is exactly a discrete summation form of \eq{contpote}, and the 
correspondence to \eq{contpote} is understood by the following 
translation rule from discrete spectrum to continuum one 
\beqa
2l\xi \quad &\ra& \quad k^2, \nn\\
\xi \quad &\ra& \quad  kdk, \label{corres} \\
\xi\sum_{l=1}^\infty \quad&\ra&\quad\integ kdk\ .\nn 
\eeqa
Remember, in the continuum expression, we have to (i) insert a step 
function $\theta(\mu-k)$ into the above momentum integration and have 
to (ii) keep $\mu<\sigma$ in order to observe a correct potential 
behavior in $\beta\ra\infty$ with finite $\mu$; namely \cite{IKM}
\beq
\lim_{\beta\ra\infty}{\tilde V}(\sigma;0,\beta,\mu) =
{\trone2\sqrt{\pi}\over(4\pi)^{D/2}\Gamma({D-1\over2})}
\int_0^\mu dk k^{D-2} (\mu-k) \ .     \label{zeropote}
\eeq
Due to the correspondence \eq{corres}, we may similarly assume an 
upper bound on the discrete momentum sum; $l<L\equiv[\mu^2/2\xi]_G$, 
where $[{\phantom G}]_G$ is the Gauss symbol, 
\beq
\lim_{\beta\ra\infty} {\tilde V}(\sigma;\xi,\beta,\mu)= 
{\trone\xi2\sqrt{\pi}\over(4\pi)^{3/2}}
\Bigl[\,{1\over2}\mu + \sum_{l=1}^L(\mu-\sqrt{2l\xi})\,\Bigr]\ . 
\eeq
This potential is still clearly a discrete form of \eq{zeropote} 
for $D=3$ under the identification of \eq{corres}.

Let us return to the generic $D$ case. The above cutting 
procedure is also valid for the generic $D$ case, because we 
naturally have a step function in 
\beqa
{\cal O}_\infty(\sigma) &\equiv& 
\lim_{\beta\ra\infty}{\cal O}_{\beta}(\sigma) \nn \\
& = & -{2\sqrt{\pi}\over\Gamma({D-1\over2})}\theta(\mu-\sigma)
\int_0^{\mu-\sigma}(t^2+2\sigma t)^{D-3\over2}dt  \\
& = & -{2\sqrt{\pi}\over\Gamma({D+1\over2})}(2\sigma)^{D-3\over2}
(\mu-\sigma)^{D-1\over2} F\Bigl({3-D\over2},{D-1\over2};{D+1\over2};
{\sigma-\mu\over2\sigma}\Bigr)\,\theta(\mu-\sigma)\ , \nn
\eeqa
where we have applied the integration formula \eq{Grad31941}. Owing 
to the presence of the step function $\theta(\mu-\sigma)$, the 
upper bound $L$ is naturally introduced here, and we figure out 
\beq
\lim_{\beta\ra\infty}{\tilde V}(\sigma;\xi,\beta,\mu) =
-{\trone\xi\over(4\pi)^{D/2}}\Bigl[\,{1\over2}{\cal O}_\infty(0)
+\sum_{l=1}^L{\cal O}_\infty(\sqrt{2l\xi})\,\Bigr]\ . \label{zeropoteD}
\eeq
This is the discrete form corresponding to \eq{zeropote}. 
The explicit forms of ${\cal O}_\infty(\sigma)$ in each $D$ are 
\beq
{\cal O}_\infty(\sigma) = {-2\sqrt{\pi}\over\Gamma({D-1\over2})}
\theta(\mu-\sigma)\times\left\{ 
\begin{array}{ll}
 \mbox{arccosh}\,(\mu/\sigma)    &\quad\mbox{for} \quad D=2  \\
 \mu - \sigma                    &\quad \mbox{for} \quad D=3 \\ 
 {1\over2}\mu\sqrt{\mu^2-\sigma^2} - {1\over2}\sigma^2
 \mbox{arccosh}\,(\mu/\sigma)    &\quad \mbox{for} \quad D=4\ , 
\end{array}\right.
\eeq
and 
\beq
{\cal O}_\infty(0) = -{2\sqrt{\pi}\over\Gamma({D-1\over2})}
{\mu^{D-2}\over D-2} \ .
\eeq
The ${\cal O}_\infty(0)$ is singular when $D=2$, however it vanishes 
under the 'physical' limit $\xi\ra0$. In Appendix B, we show how the 
discrete form \eq{zeropoteD} reproduces the continuum one 
\eq{zeropote} in the limit $\xi\ra0$ via \eq{corres}. 

Now let us examine the phase structure on the $\xi$-$\mu$ plane. It is 
enough to analyze the 1'st order critical equation defined by 
\eq{cri1order1} and \eq{cri1order2}. The reason will become clear at the 
bottom of this paragraph. Since the pure thermal part 
\eq{zeropoteD} does not depend on $\sigma$ any more, the gap 
equation is the same as the one for the zero temperature potential 
$V(\sigma;\xi)$,
\beq
0={1\over\lambda_R}+\trone\integ{ds\over(4\pi s)^{D/2}}
s\xi\coth(s\xi)\Bigl[e^{-s}(1-2s)-e^{-sm^2}\Bigr]. \label{zerogap}
\eeq
(Of course, this can be derived from \eq{gap} with the use 
of \eq{zeroT}.) Eliminating the renormalized coupling from \eq{vs} 
with using \eq{zerogap}, and applying the following 
formula~\footnote{
This formula is derived from the expansion \eq{coth} and the 
definition of zeta function.}
\beq
\integ ds s^{\mu-1}e^{-\beta s}\coth s=\Gamma(\mu)\Bigl[2^{1-\mu} 
\zeta(\mu,1+{\beta\over2}) + \beta^{-\mu}\Bigr], \qquad \mbox{Re}\,\mu>1 
\label{formula1}
\eeq
$V(\sigma;\xi)|_{\sigma=m}$ turns out to be 
\beqa
V(m;\xi)&=&{\trone2^{D/2}\xi^{{D\over2}-1}\over2(4\pi)^{D/2}}
\Gamma(1-{D\over2})\left[ {1\over2}m^2(1-{D\over2})
\zeta\Bigl(2-{D\over2},1+{m^2\over2\xi}\Bigr)\right. \nn\\ 
&+&\left.\xi\zeta\Bigl(1-{D\over2},1+{m^2\over2\xi}\Bigr)
-\xi\zeta\Bigl(1-{D\over2}\Bigr)
+{m^2\over4}(2-{D\over2})({2\xi\over m^2})^{2-D/2}\right]. \label{vms}
\eeqa
Combining this and \eq{zeropoteD}, we obtain the 1'st order 
critical equation on the $\xi$-$\mu$ plane
\beqa
0&=&{-2\over\Gamma(1-{D\over2})} (2\xi)^{2-{D\over2}}
\Bigl[\,{1\over2}{\cal O}_\infty(0) + 
\sum_{l=1}^L{\cal O}_\infty(\sqrt{2l\xi}\,) \Bigr] \nn\\
&+& m^2 (2-D)\zeta\Bigl(2-D,1+{m^2\over2\xi}\Bigr) 
    + {4-D\over2}(2\xi)^{2-{D\over2}} m^{D-2} \nn\\
&+& 4\xi\zeta\Bigl(1-{D\over2},1+{m^2\over2\xi}\Bigr) 
    -4\xi\zeta\Bigl(1-{D\over2}\Bigr)\ ,   \label{ximu1cri}
\eeqa
where $m$ should be determined from \eq{zerogap}. (For example, we 
can take $m=1$ as a solution if we choose, instead of \eq{coupling}, 
\beq
{1\over\lambda_R}=\trone\integ{ds\over(4\pi s)^{D/2}}\,s\xi\coth(s\xi)
\cdot 2se^{-s}.
\eeq
This coupling choice makes our critical equations much simpler, 
however it is not essential for the numerical analyses of phase 
structures discussed later). Under the choice \eq{coupling}, the 
mass~\footnote{For example, a solution is $m=1.02107$ for 
$\xi=0.3$ in $D=3$.} is determined by 
\beqa
0&=&2-\xi(3-D)-\xi m^{D-4} 
-(2\xi)^{{D\over2}-1}\zeta\Bigl(2-{D\over2},1+{m^2\over2\xi}\Bigr) \nn\\
&& + (2\xi)^{{D\over2}-1}\zeta\Bigl(2-{D\over2},1+{1\over2\xi}\Bigr) 
-(2\xi)^{{D\over2}-2}(4-D)\zeta\Bigl(3-{D\over2},1+{1\over2\xi}\Bigr).
\label{gapm}
\eeqa
The dynamical mass determined by \eq{gapm} grows as the strength $\xi$ 
of the external magnetic field becoming large (see Fig.1), whose feature 
coincides with the one argued in \cite{Naf}. Note that we can not take the 
limit $m\ra0$ without circumventing divergence for $D<4$ in this equation. 
This means there is no 2'nd order phase transition on the $\xi$-$\mu$ plane. 

Substituting the solutions of \eq{gapm} into \eq{ximu1cri}, we find 1'st 
order transitions (see Fig.2) as the solution of \eq{ximu1cri}. The equation \eq{ximu1cri} can be shown to coincide with the following equation \cite{IKM} 
in the limit $\xi\ra0$ (The proof is in Appendix C), 
\beq
\left({\mu\over m}\right)^D = {D-1\over2}\Bigl(1-{D\over2}\Bigr)
{\rm B}\Bigl(1-{D\over2},{D-1\over2}\Bigr),  \label{c1line}
\eeq
which has the solutions $\mu=m/\sqrt{2}$ for $D=2$ and 
$\mu=m$ for $D=3$ \cite{Wol,CM}. 

\vspace{4mm}
\begin{minipage}[t]{7cm}  
\epsfxsize=8.5cm
\epsfbox{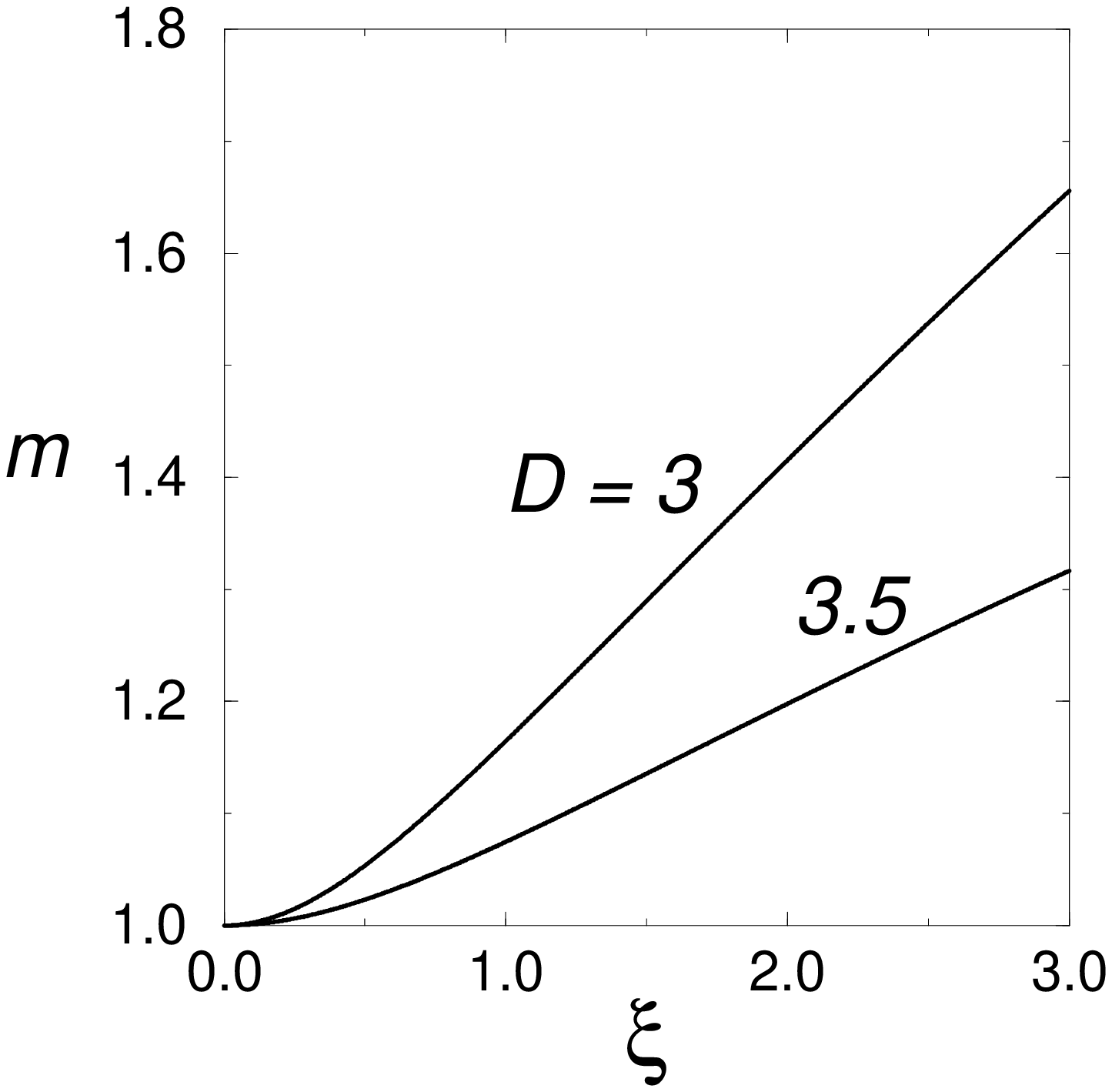}
  {\bf Fig. 1:}   The dynamical mass $m$ for magnetic dominant cases 
                    on  $T=0$ plane.   
\end{minipage}
\hspace{0cm}
\begin{minipage}[t]{7cm}
\epsfxsize=8.5cm
\epsfbox{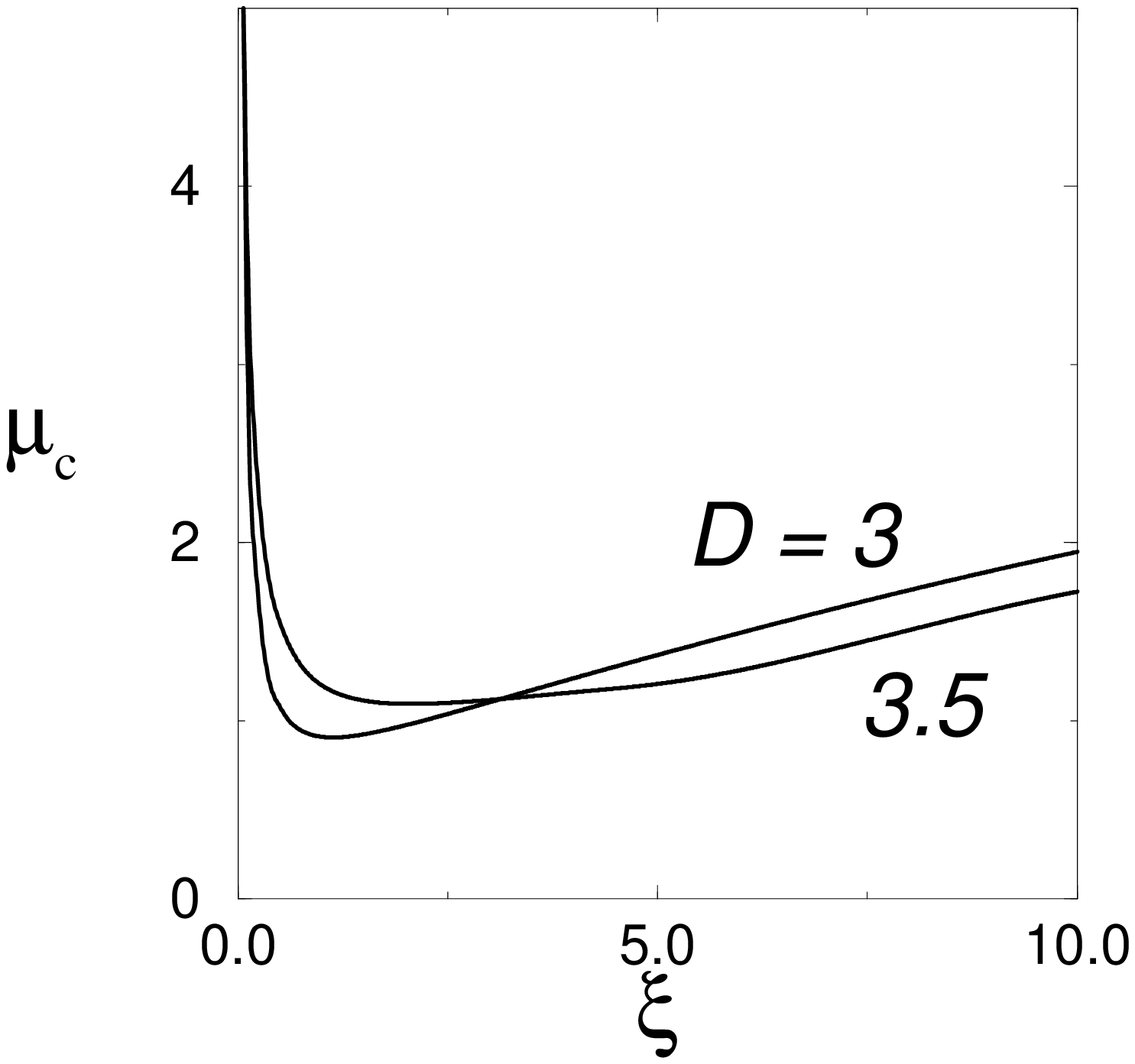}
  {\bf Fig. 2:}    The values of critical chemical potential $\mu_c$ 
                     on the $T=0$ plane 
                     as a function of the 
                     external magnetic field $\xi$. 
\end{minipage}

\subsection{The phase structures}
\indent


In Fig.3, we illustrate several critical curves which represent 
various sections of the second order critical surface in the 
$\xi$-$T$-$\mu$ space for $D=3$. In this case, the point 
$(\xi,T,\mu)=(0,0,1)$ is the unique tricritical point, since 
\eq{tricripoint} has the solution $(\beta\mu)^{-1}=0$, which means 
$T=0$ (for $\mu\not=0$). Fig.4 is the case of $D=4-\epsilon$, where 
we take $\epsilon=0.5$. No tricritical point appears in this case. 
We could not let $\epsilon$ to be much smaller than this value due 
to a singular situation of numerical computations. However we can 
easily read some signs of the magnetic catalyst and the disappearance 
of the tricritical point, if we extrapolate from the following  
behaviors of larger values $\epsilon=1.5$ and $2.0$. 

First, let us observe the $D=2.5$ ($\epsilon=1.5$) diagrams in Fig.5, 
where the tricritical point starts appearing around $0.75<\mu<0.76$. 
These tricritical points are determined from intersections between 
two curves solved from \eq{j} and \eq{m} like the way shown in Fig.6. 
Next, let us watch the $D=2$ diagrams (Fig.7). The tricritical line 
(points) and first order transitions are observed for $0.61<\mu$. 
The phase diagram is very similar to Fig.5, and these diagrams 
exhibit that strong $\xi$ makes phase broken (magnetic catalyst) 
even in these unphysical dimensions $2\leq D<3$. Now we gather the 
tricritical points of each dimension in Fig.8. It clearly shows that 
the tricritical line moves toward a lower temperature region as $D$ 
increasing. And $D=3$ is the terminal case that the tricritical 
point appears. We therefore deduce that there is no tricritical 
point in $D=4-\epsilon$ for small $\epsilon$ whilst observing the 
magnetic catalyst in finite $T$ and $\mu$.

Let us summarize the phase structures. We have observed that the 
critical surfaces in $D=3$ and $4-\epsilon$ are of second order, 
and magnetic fields tend to make the chiral symmetry broken even at 
finite $T$ and $\mu$. The large values of $\mu$ tend to revive the 
chiral symmetry, and the critical $\mu$ grows as $\xi$ increasing 
(for all $T$). There are first order transitions not at finite 
$T$ but at zero temperature shown in Fig.2, where the $\xi=0$ axis 
is singular and discontinuous. 

\vspace{1cm}
\begin{minipage}[t]{7cm}  
\epsfxsize=8.5cm
\epsfbox{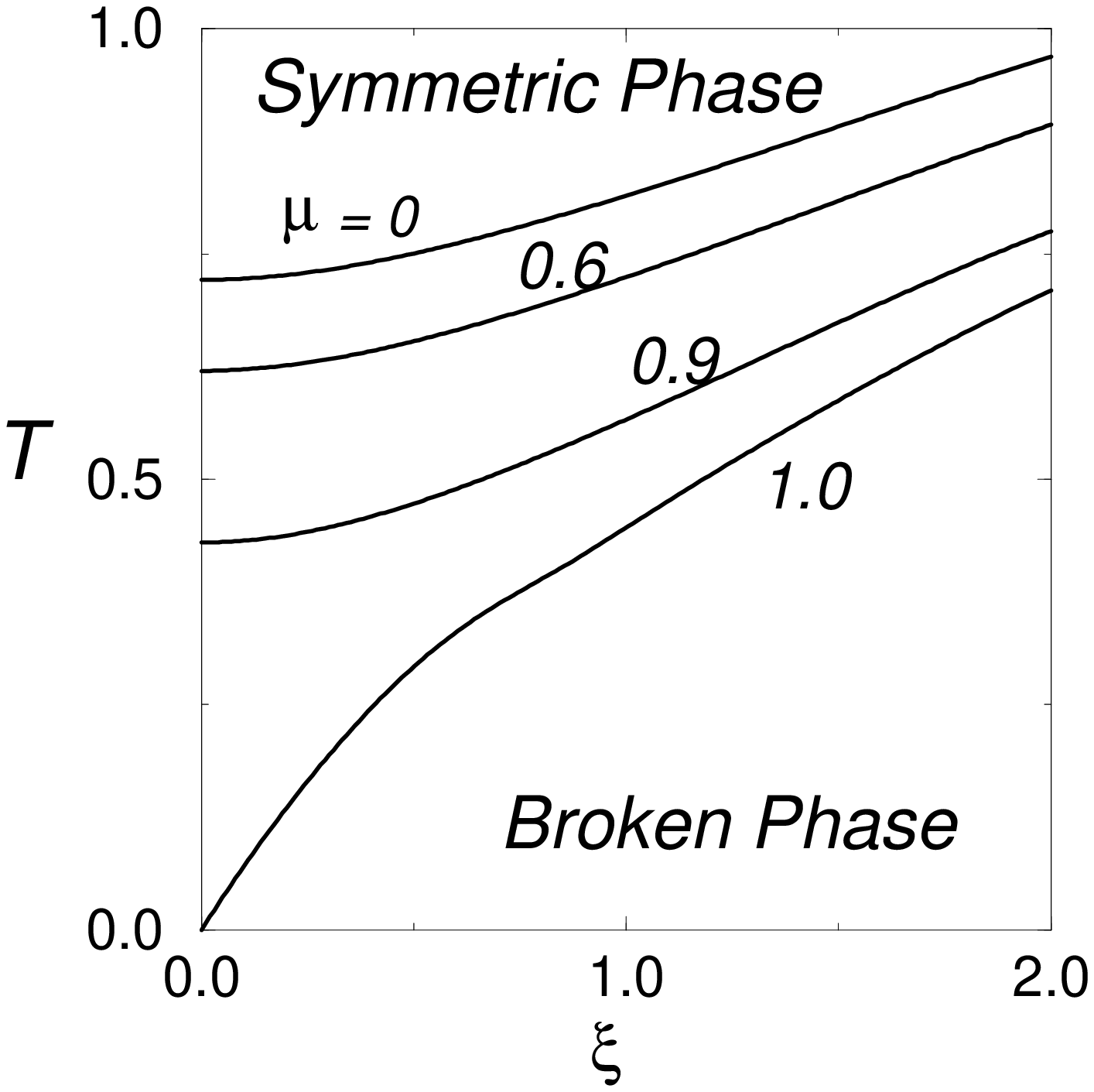}
  {\bf Fig. 3.} The critical lines for magnetic dominant cases in $D=3$.
\end{minipage}
\hspace{0cm}
\begin{minipage}[t]{7cm}
\epsfxsize=8.5cm
\epsfbox{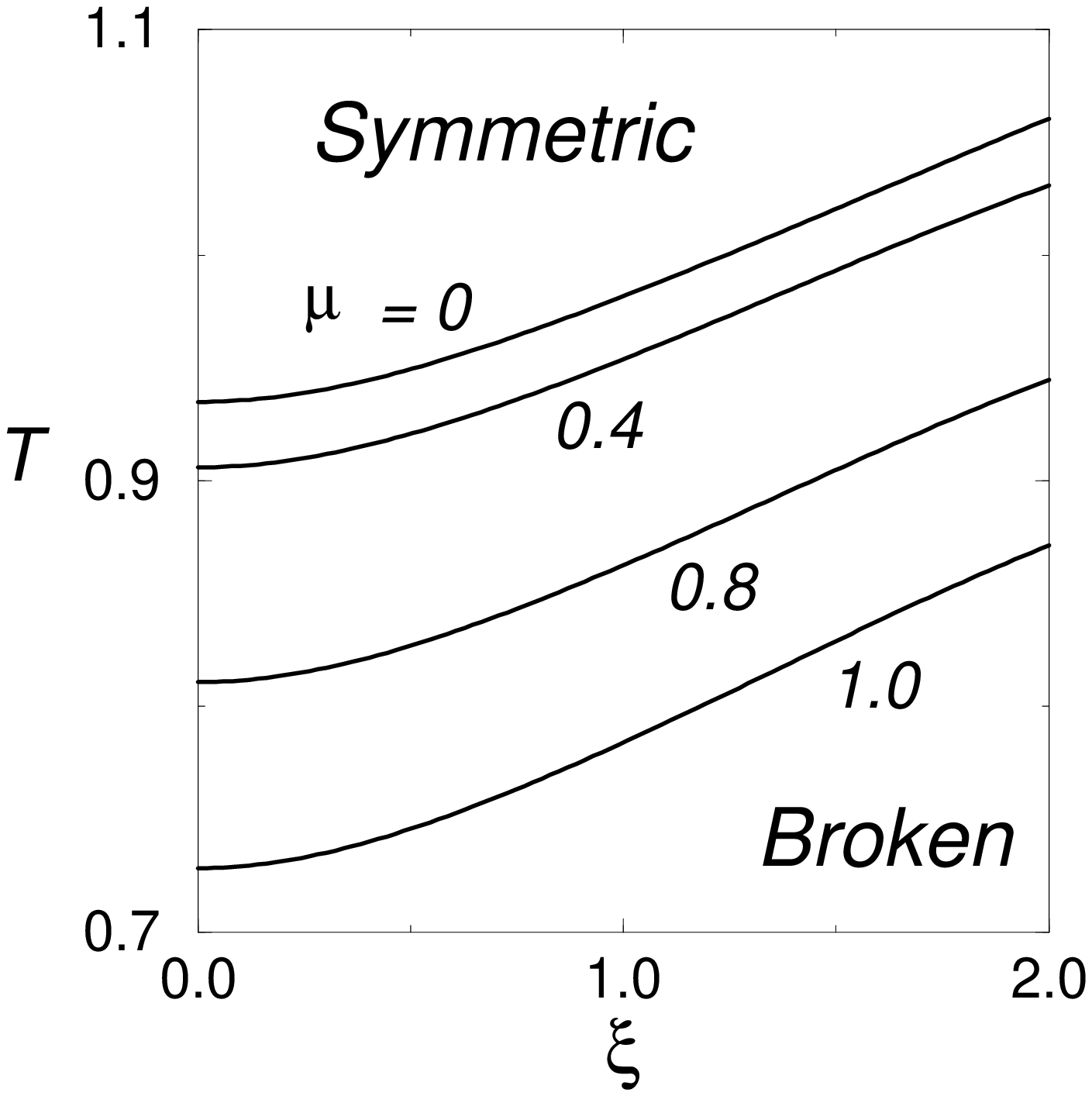}
  {\bf Fig. 4:}  The critical lines for magnetic dominant cases in 
                 $D=4 - \epsilon$, where $\epsilon$ is chosen to be $0.5$. 
\end{minipage}
\vspace{5mm}

\vspace{1cm}
\begin{minipage}[t]{7cm}  
\epsfxsize=8.5cm
\epsfbox{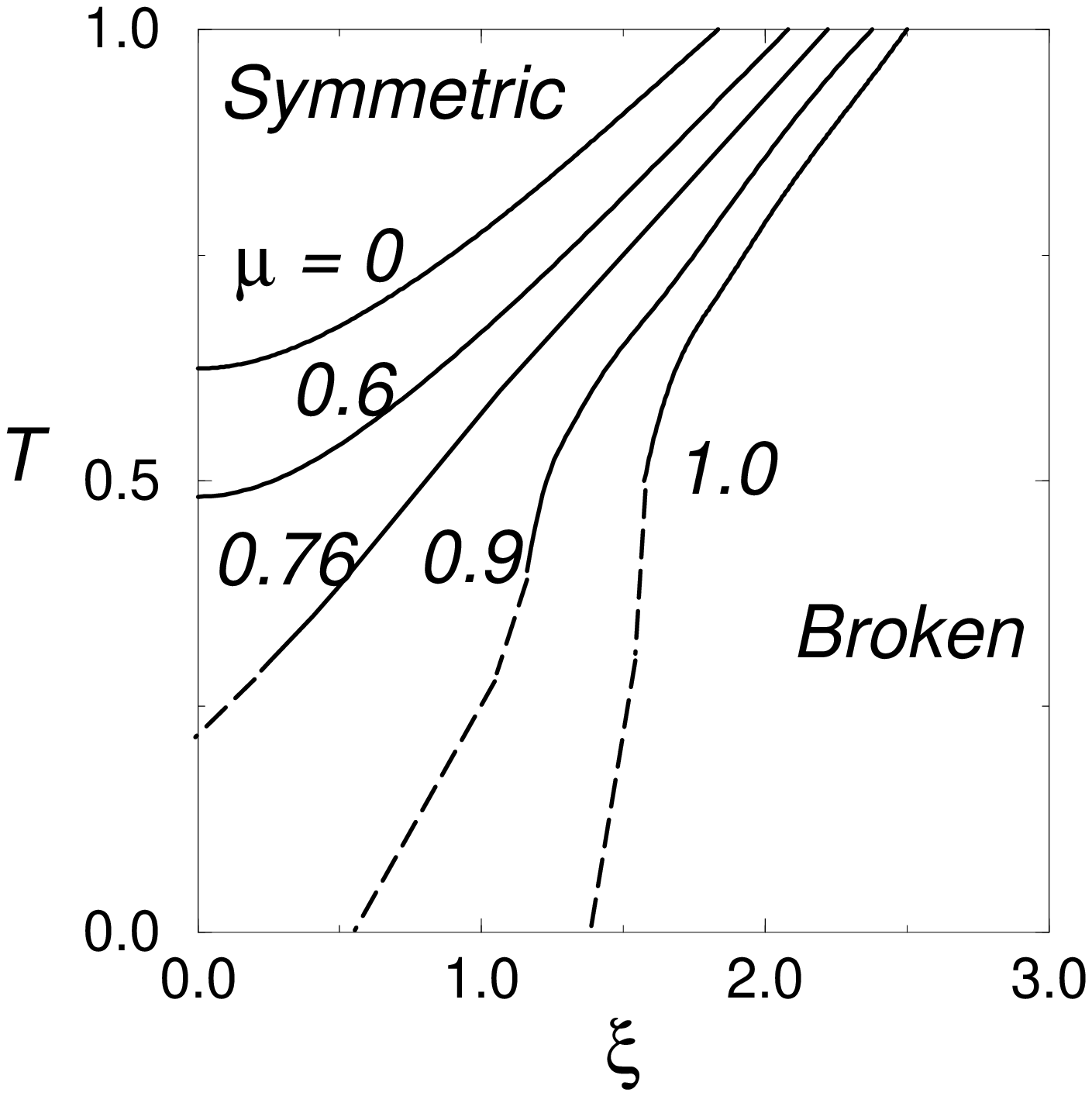}
  {\bf Fig. 5:}  The critical lines for magnetic dominant cases in $D=2.5$.
\end{minipage}
\hspace{0cm}
\begin{minipage}[t]{7cm}
\epsfxsize=8.5cm
\epsfbox{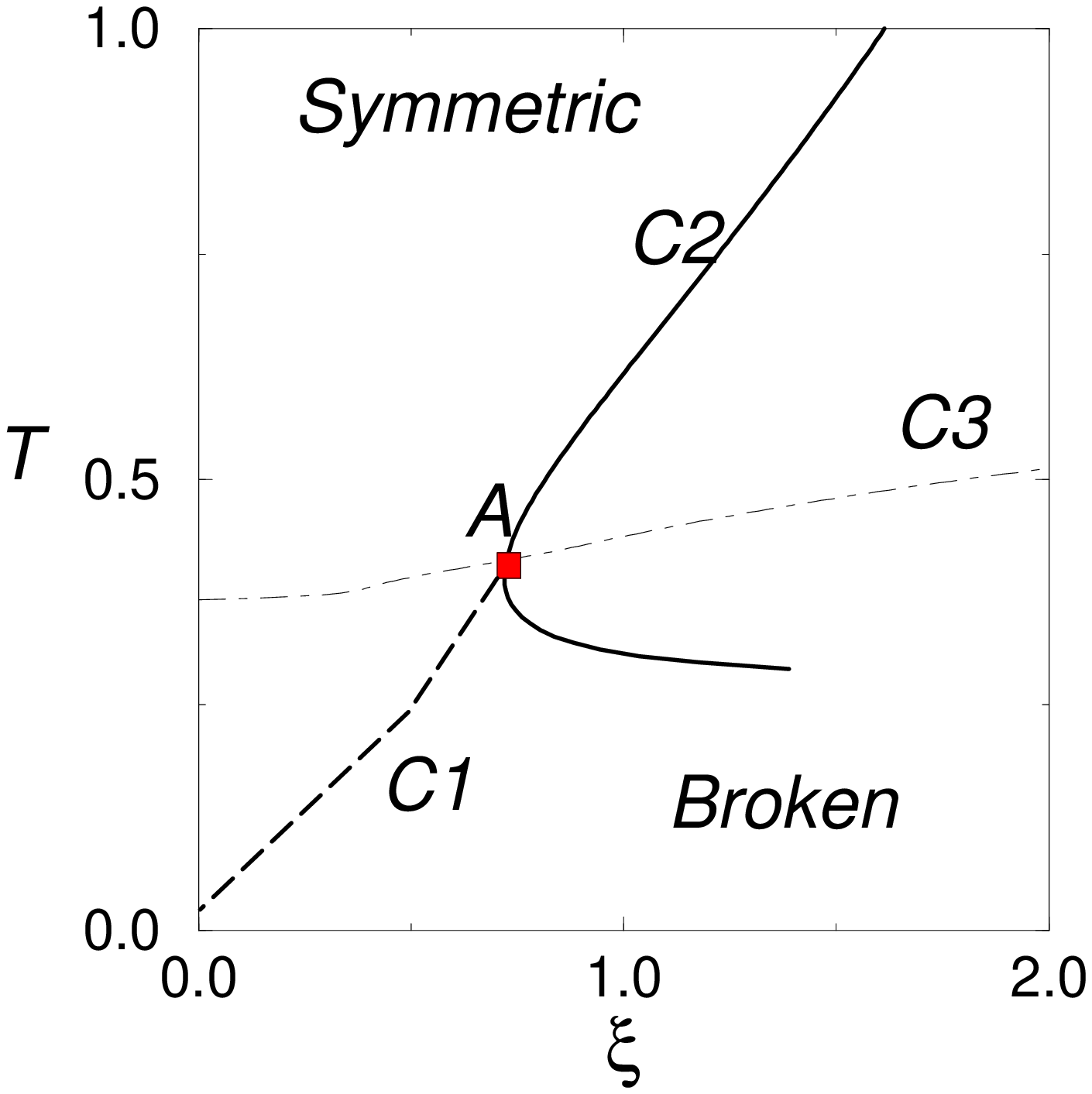}
  {\bf Fig. 6:}  How to determine the tricritical point $A$. This 
example is the case of $\mu=0.7$, $D=2$ (magnetic dominant). 
       The $C2$ and $C3$ are the solutions of eqs. \eq{j} and \eq{m}. 
       The $C1$ represents the solution of \eq{c1equation}.
\end{minipage}
\vspace{5mm}
\vspace{1cm}
\begin{minipage}[t]{7cm}  
\epsfxsize=8.5cm
\epsfbox{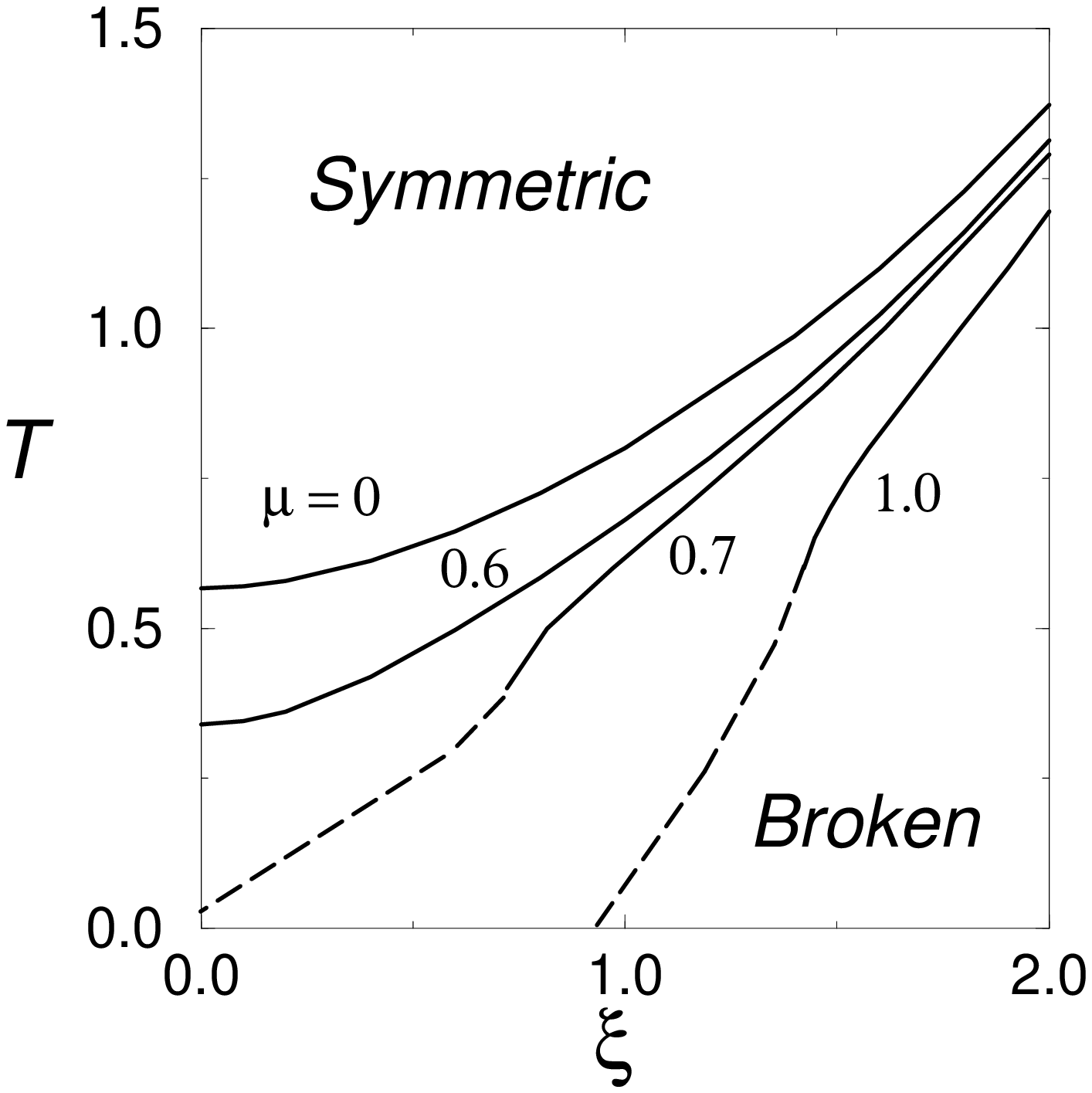}
  {\bf Fig. 7:} The critical lines in $D=2$ 
                  for magnetic-like external field. 
                  The $T=0$ plane is singular. 
\end{minipage}
\hspace{0cm}
\begin{minipage}[t]{7cm}
\epsfxsize=8.5cm
\epsfbox{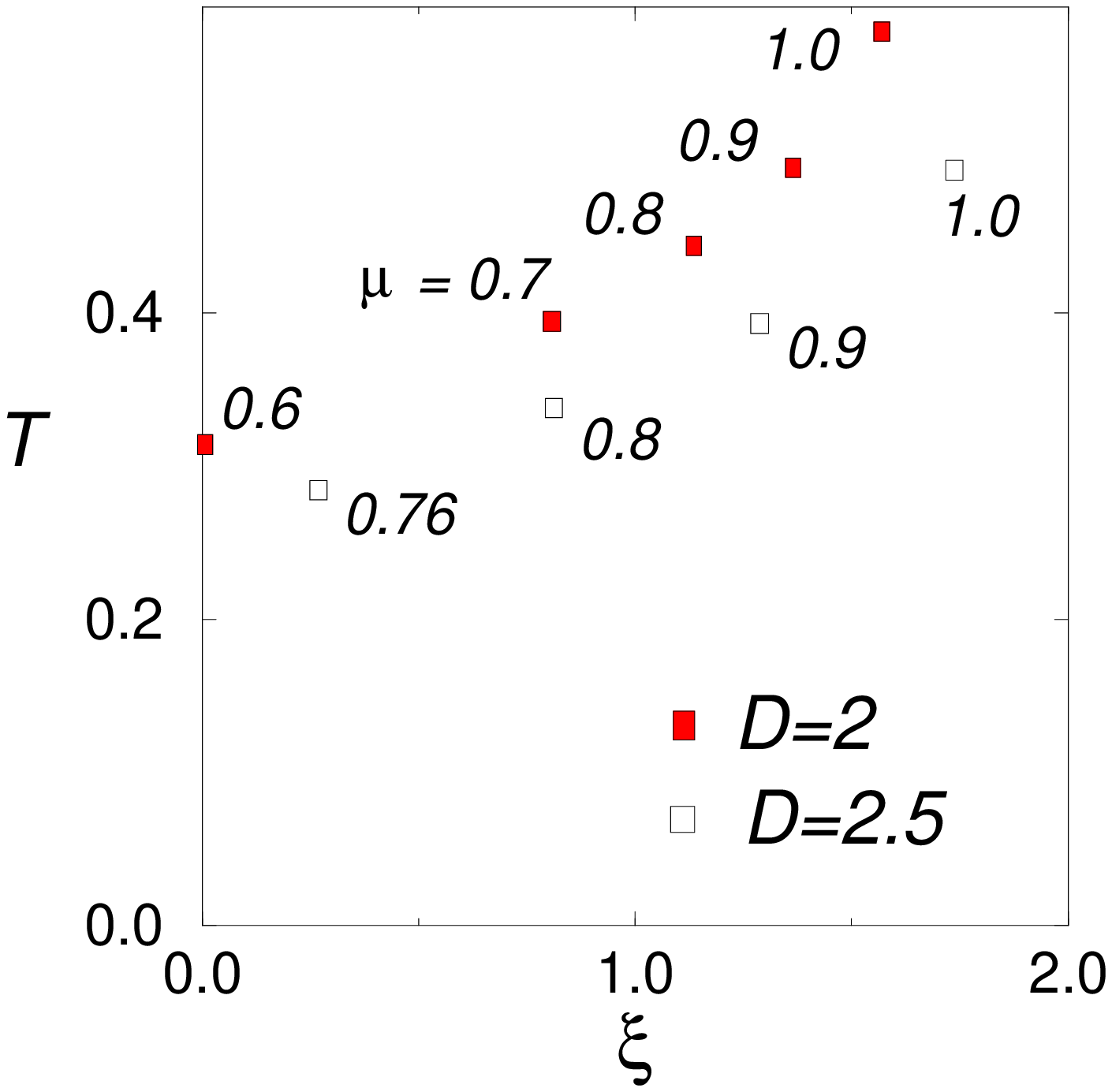}
  {\bf Fig. 8:} The tricritical points projected on the $T-\xi$ plane 
     for magnetic dominant cases. 
\end{minipage}
\vspace{5mm}

\section{The electric dominant cases}
\subsection{The imaginary part of effective potential}
\setcounter{equation}{0}
\indent

%
In contrast with the $B$-dominant cases, the effective potential 
\eq{b} gives rise to an imaginary part due to the instability of 
fermion vacuum in the $E$-dominant cases \cite{KKM,Stone,PT}. 
In fact, the function $s\xi {\rm coth} (s \xi)$ in the integrand 
of \eq{b} is now $s\rho \cot (s\rho)$ with changing $\xi=i\rho$, and 
the integrand then contains an infinite number of the singular 
points $s=n\pi/\rho,(n=1,2,\cdot\cdot\cdot)$ accordingly. 
Then we must displace the contour of integration 
($-\infty < s < \infty$) by $i \epsilon $ along the imaginary axis 
of the complex $s$-plane. The contribution of each singular point is 
taken into account by making use of the $i\epsilon$ rule (for example, 
\cite{Stone})
\begin{eqnarray}
  \fr{1}{s + i \epsilon - \fr{n \pi}{\rho}} = 
\fr{{\rm P}}{ s - \fr{n \pi}{\rho}} - i \pi \delta 
\left( s - \fr{n \pi}{\rho} \right),
\end{eqnarray}
where P stands for the principal-value symbol. By virtue of this rule, 
the imaginary part of the effective potential consists of 
\beqa
{\rm Im}\, V_{\beta,\mu}(\sigma;\rho)&=& -\trone
{\rho^{(D-1)/2}\over2^D\pi^{D-2}\beta}\sum_{n=1}^\infty
n^{1-D\over2}e^{-n\pi(\sigma^2-\mu^2)/\rho}
\Theta_2({2n\pi\mu\over\rho\beta},i{4n\pi^2\over\rho\beta^2}) \nn\\
&=&-\trone{\rho^{(D-1)/2}\over2^{D-1}\pi^{D-2}\beta}
\mbox{Re}\,\sum_{m=0}^\infty\exp\Bigl[-{\pi\over\rho}
\{\sigma^2+(\pi{2m+1\over\beta}+i\mu)^2\,\}\Bigr]\nn\\
&&\ \times\Phi\Bigl({D-1\over2};{i\over2\rho}
\{\sigma^2+(\pi({2m+1\over\beta}+i\mu)^2\,\},\,1\Bigr)\ ,
\label{impo}
\eeqa
where $\Phi(z;s,a)$ is the Lerch transcendental function 
defined by \eq{lerch}. This can be regarded as a probability, per 
unit space-time volume, for a fermion pair-production from the 
vacuum by an external $E$ field. Taking the $T=\mu=0$ limit
\beq
\lim_{T,\mu\ra0}{\rm Im}\,V_{\beta,\mu}(\sigma;\rho)=
-{\trone\rho^{D/2}\over2^{D+1}\pi^{D-1}}e^{-\pi\sigma^2/\rho}
\Phi\Bigl({D\over2};i{\sigma^2\over2\rho},1\,\Bigr)\ ,
\label{zeroimV}
\eeq
we reproduce the known results \cite{KKM,Kli3}
\beq
{\rm Im}\,V(\sigma) = \left\{ 
\begin{array}{ll}
 \trone\rho/(8\pi)\cdot\ln(1-e^{-\sigma^2\pi/\rho}) 
                               &\quad\mbox{for} \quad D=2  \\
 - \trone\rho^{3/2}/(4\pi)^2\cdot e^{-\pi\sigma^2/\rho}
             \Phi({3\over2};i{\sigma^2\over2\rho},1)
                               &\quad \mbox{for} \quad D=3. 
\end{array}\right.  \label{sample}
\eeq
One can rewrite \eq{zeroimV} in the following form using the Lerch 
transformation \eq{Ltrans},
\beqa
{\rm Im}\,V(\sigma)&=&
- {\trone\rho^{D/2}\over4(2\pi)^{D/2}}\Gamma(1-{D\over2})
\left[ \exp\Bigl[{\pi\over2}i(1-{D\over2})\Bigr]
\zeta\Bigl(1-{D\over2},i{\sigma^2\over2\rho}\Bigr)\right. \nn\\
&&\  + \left.\exp\Bigl[-{\pi\over2}i(1-{D\over2})\Bigr]
\zeta\Bigl(1-{D\over2},1-i{\sigma^2\over2\rho}\Bigr) \right]\ .
\eeqa
We shall search for the lowest energy configuration of $\sigma$ 
solving the critical surface equations derived in sect.2. This ground 
state solution $\sigma(E)$ minimizes the real part of the effective 
potential, and the chiral symmetry is broken accordingly. On the 
contrary, we expect that a strong external field might induce a 
restoration of chiral symmetry. However, we can not conclude 
immediately whether or not the chiral symmetry is restored even if 
the fermion mass vanishes, because the effective potential possesses 
the imaginary part that represents a decay of "symmetry vacuum" to a 
fermion pair. Indeed, in the case of $T=\mu=0$, the chiral mass term 
$\sigma(E)\langle{\bar\psi}\psi\rangle$ does not vanish even when 
$\sigma(E)=0$ in $D=2$ \cite{KKM}. 

Hence we determine the symmetry restoration by the criterion whether 
the chiral mass term vanishes or not ~\cite{KKM,Kli2}
\beq
\lim_{\rho\ra\rho_c}
 \sigma(\rho) {\rm Im}\,V'(\sigma(\rho);\rho,\beta,\mu) =0,
\label{kkmcri}
\eeq
where we have applied the fact that $\langle{\bar\psi}\psi\rangle$ is 
related to the ${\rm Im}\,V'$ by
\beq
\langle{\bar\psi}\psi\rangle_{\sigma=\sigma(\rho)} 
= iN{\rm Im}\,V'(\sigma(\rho);\rho,\beta,\mu) 
- {N\over\lambda} \sigma(\rho), 
\eeq
and $\sigma(\rho)$ denotes a solution of gap equation, i.e., 
${\rm Re}\,V'=0$. To check this criterion, it is sufficient to see 
whether or not the ${\rm Im}\,\,V'(\sigma;\rho,\beta,\mu)$ is 
finite as $\sigma\ra0$. From \eq{impo}, we have
\beq
{\rm Im}\, V'(\sigma;\rho,\beta,\mu)= \sigma\trone
{\rho^{(D-3)/2}\over2^{D-1}\pi^{D-3}\beta}\sum_{n=1}^\infty
n^{3-D\over2}e^{-{n\pi\over\rho}(\sigma^2-\mu^2)}
\Theta_2({2n\pi\mu\over\rho\beta},i{4n\pi^2\over\rho\beta^2}).
\label{impod}
\eeq
The upper bound for the summation value in \eq{impod} can be 
estimated in the following way
\beqa
& &\sum_{n=1}^\infty
n^{3-D\over2}e^{-{n\pi\over\rho}(\sigma^2-\mu^2)}
\Theta_2\Bigl({2n\pi\mu\over\rho\beta},
i{4n\pi^2\over\rho\beta^2}\Bigr)\nn\\
&=& \sum_{n=1}^\infty n^{3-D\over2}\sum_{m=-\infty}^\infty
\exp\Bigl[-n{\pi\over\rho} 
\{\sigma^2+(\pi{2m+1\over\beta}-i\mu)^2\,\}\Bigr] \nn\\
&\leq& \integ dx x^{3-D\over2}\sum_{m=-\infty}^\infty
\exp\Bigl[-x{\pi\over\rho} 
\{\sigma^2+(\pi{2m+1\over\beta}-i\mu)^2\,\}\Bigr].
\label{inequality}
\eeqa
Performing the $x$-integration 
\beq
{\rm RHS}\,\,\eq{inequality} = 
\Gamma({5-D\over2})\left({\rho\over\pi}\right)^{5-D\over2} 
\sum_{m=-\infty}^\infty 
\Bigl[ \sigma^2+(\pi{2m+1\over\beta}-i\mu)^2 \Bigr]^{D-5\over2},
\eeq
we find it finite as $\sigma\ra0$ 
\beq
\lim_{\sigma\ra0}{\rm RHS}\,\,\eq{inequality} = 
2\Gamma({5-D\over2}) 
\left({2\pi^2\over\rho\beta}\right)^{D-5\over2} {\rm Re}\,
\zeta\Bigl(5-D,{1\over2}+i{\beta\mu\over2\pi}\Bigr)\,.\label{impolim}
\eeq
The criterion is therefore satisfied
\beq
\lim_{\sigma\ra0}\sigma{\rm Im}\,V'(\sigma;\rho,\beta,\mu)=0,
\label{discri2}
\eeq
and we conclude that the chiral symmetry is restored for finite 
$T$ and $\mu$ in $2\leq D<4$. 

Note that RHS of \eq{impolim} vanishes right at a tricritical point, 
whose necessary condition is \eq{tricripoint}. This is not the case 
with 2'nd order transitions and suggests that the multi-criticality 
can be mightier than the vacuum instability. As a by-product,
we obtain a necessary condition for the tri-criticality on the 
imaginary part
\beq
\lim_{\sigma\ra0}{1\over\sigma}{\rm Im}\, 
V'(\sigma;\rho,\beta,\mu)=0,             \label{discri3}
\eeq
and it leads to 
\beq
{1\over\beta} \sum_{n=1}^\infty
n^{3-D\over2}e^{n\pi\mu^2\over\rho}
\Theta_2\Bigl({2n\pi\mu\over\rho\beta},
i{4n\pi^2\over\rho\beta^2}\Bigr)=0\,.
\eeq
It is interesting to note that the powers of $\sigma$ in front of 
${\rm Im}\,V'$ in \eq{discri2} and \eq{discri3} differ by two from 
each other. Hence \eq{discri2} and \eq{discri3} may be called a kind 
of discrete versions of \eq{cond1} and \eq{cond2}. 

In the zero temperature cases, one may derive exact expressions 
for ${\rm Im}\,V'$ from \eq{impo} by taking the $T$,$\mu\ra0$ limit 
like done in \eq{sample}. Here, we only put a brief statement about 
their power dependence on $\sigma$ in this limit. Using 
\beq
\lim_{\beta\ra\infty}{1\over\beta}\sum_{m=-\infty}^\infty
\exp\Bigl[-x{\pi\over\rho}\{\sigma^2+(\pi{2m+1\over\beta})^2\,\}\Bigr] 
= e^{-x{\pi\over\rho}\sigma^2} {1\over\sqrt{4\pi x}}\,,
\eeq
we obtain 
\beq
\lim_{\beta\ra\infty}{1\over\beta}
{\rm RHS}\,\,\eq{inequality} = {1\over2}\pi^{{D\over2}-3}
\rho^{5-D\over2} \Gamma(2-{D\over2}) \sigma^{D-4}\,,  \label{impos}
\eeq
and its power dependence on $\sigma$ exactly coincides with 
known results \cite{KKM,Kli2} 
\beq
\lim_{T,\mu\ra0} {\rm Im}\,V'(\sigma;\rho,\beta,\mu)\sim \left\{ 
\begin{array}{ll}
 \sigma^{-1}    &\quad\mbox{for} \quad D=2,  \\
 {\rm const.}   &\quad \mbox{for} \quad D=3, \\ 
\sigma^{1-\epsilon} &\quad \mbox{for} \quad D=4-\epsilon. \\ 
\end{array}\right.               \label{impozero}
\eeq
Namely, chiral symmetry is restored in $D=3$ \cite{Kli2} and 
$4-\epsilon$, and broken in $D=2$ \cite{KKM}.

\subsection{The duality and phase structures}
\indent

In this subsection, we present phase diagrams with analyzing the real 
parts of the critical equations \eq{j} and \eq{m}. Also, we point out 
a dual relation between the electric and magnetic dominant effective 
potentials (at $T=0$) under the rotation $\xi\ra i\rho$. We follow the 
same method employed in \cite{KKM,Kli2} in order to evaluate the real 
parts (principal values). 

Let us introduce the following compact notations for the real part 
effective potential \eq{repotential} (to avoid lengthy expressions),
\beq
{\rm Re}\,V_R(\sigma;\rho,\beta,\mu) = 
{1\over2\lambda_R}\sigma^2 + 
{\trone\over2(4\pi)^{D/2}}\Bigl(I_1(\sigma)-I_1(0)\Bigr)\ ,
\label{RVR}
\eeq
where
\beq
I_1(\sigma)={\rm P}\,\integ \fr{ds}{s^{D/2}} s\rho\cot s\rho\cdot 
f(s,\beta,\mu)\ ,  \label{I1}
\eeq
and
\beq
f(s, \beta, \mu) =  {1\over s} e^{-s\sigma^2} e^{s \mu^2} 
\fr{\sqrt{4\pi s}}{\beta} \teta + \sigma^2 e^{-s}(1 - 2s). 
\label{K1}
\eeq
Furthermore, split $I_1(\sigma)=A(\sigma)+B(\sigma)$ into the following 
two pieces for convenience:
\begin{eqnarray}
A(\sigma)&=& \rho^{D/2-1}\intep \fr{d \tau}{(4\pi)^{D/2}} \tau^{1-D/2} 
     \cot{\tau}\cdot f(\fr{\tau}{\rho}, \beta, \mu) \label{I1a} \\
B(\sigma)&=& \rho^{D/2-1} \sum_{n=1}^{\infty} {\rm P} 
   \int_{n\pi-\pi/2}^{n\pi+\pi/2} \fr{d \tau}{(4\pi)^{D/2}} \tau^{1-D/2} 
     \cot{\tau}\cdot f(\fr{\tau}{\rho}, \beta, \mu), \nn\\
&=& \rho^{D/2-1} \sum_{n=1}^{\infty} \int_0^{\pi/2} 
    \fr{d \tau}{(4\pi)^{D/2}} \cot{\tau}\cdot 
\Bigl[ 
  \fr{f(\fr{\tau+n\pi}{\rho},\beta,\mu)}{(\tau+n\pi)^{D/2-1}} 
  - (\tau\ra -\tau)  \Bigr]\ .  \label{I1b}
\end{eqnarray}
These quantities $A$ and $B$ have another expressions. Namely, 
the integration in \eq{I1a} can be performed 
\beqa
A(\sigma)&=&-2\sum_{k=0}^{\infty}\zeta(2k)
\left({\rho\over\pi}\right)^{2k}
\left[2{\sqrt{4\pi}\over\beta} \mbox{Re}\,\sum_{m=0}^\infty
\Bigl[\sigma^2+(\pi{2m+1\over\beta}+i\mu)^2\Bigr]
^{-2k+{D-1\over2}}\right.\nn\\
&&\quad \times \gamma\Bigl({1-D\over2}+2k,
{\pi\over2\rho}\{\sigma^2+(\pi{2m+1\over\beta}+i\mu)^2\}\Bigr) \\
&&\quad +\left.\sigma^2\gamma(1-{D\over2}+2k,{\pi\over2\rho})
-2\sigma^2\gamma(2-{D\over2}+2k,{\pi\over2\rho}) \nn\right]\ .
\eeqa
Here we have applied the formula
\beq
\int_0^{\pi/2}dx x^ze^{-\alpha x}\cot x =
-2\sum_{k=0}^\infty {\zeta(2k)\over\pi^{2k}}\alpha^{-2k-z}
\gamma\Bigl(z+2k,{\pi\over2}\alpha\Bigr)\ ,
\eeq
where $\zeta(z)$ is the zeta function, and $\gamma(z,p)$ the 
incomplete gamma function defined by
\beq
\gamma(z,p)=\int_0^p e^{-t}t^{z-1}dt\qquad\mbox{Re}\,z>0\ .
\eeq
The summation of the second line in \eq{I1b} can be performed
\beqa
B(\sigma)&=&
\int_0^{\pi/2}d\tau\cot\tau\cdot\left[
{\sqrt{4\pi}\over\beta}2\mbox{Re}\,\sum_{m=0}^\infty
\exp\Bigl[-{\tau+\pi\over\rho}\{\sigma^2+ 
(\pi{2m+1\over\beta}+i\mu)^2\,\}\Bigr]\right.  \nn\\
&&\quad\times\left({\rho\over\pi}\right)^{D-1\over2}
\Phi\Bigl({D-1\over2};{i\over2\rho}
\{\sigma^2+(\pi{2m+1\over\beta}+i\mu)^2\,\},1+{\tau\over\pi}\Bigr) \nn\\
&&\quad+\sigma^2\left({\rho\over\pi}\right)^{{D\over2}-1}
e^{-(\tau+\pi)/\rho}
\Phi({D\over2}-1;{i\over2\rho},1+{\tau\over\pi}) \nn\\
&&\quad-\left.2\sigma^2\left({\rho\over\pi}\right)^{{D\over2}-2}
e^{-(\tau+\pi)/\rho}
\Phi({D\over2}-2;{i\over2\rho},1+{\tau\over\pi}) 
-(\tau\ra -\tau) \right] \ .
\eeqa
Similarly, we rewrite the second and third order critical equations 
\eq{j} and \eq{m} in a straightforward way,
\begin{eqnarray}
I_2(\sigma)+2\Gamma(2-{D\over2})=0,   \label{criticalE}
\end{eqnarray}
\beq
       I_3(\sigma) =0,
\eeq
where the explicit forms of $I_2$ and $I_3$ are given in Appendix D. 

For simplicity, let us begin with analyzing the $T=\mu=0$ case. 
In this case, \eq{K1} simplifies because of \eq{zeroT} (If one wants 
to consider the $\mu\not=0$ situation, it is necessary to follow the 
similar procedure analyzed in sect.3, and see also \cite{CD}). 
Let us write down $I_1$ and $I_2$: 
\beqa
I_1(\sigma)&=&\rho^{D/2-1}\int_0^{\pi/2}d\tau\cot\tau\cdot\left[
\rho\tau^{-D/2} e^{-\tau\sigma^2/\rho} +
\sigma^2\tau^{1-D/2}e^{-\tau/\rho}(1-{2\tau\over\rho})\right. \nn\\
&&+ \pi^{-D/2}\Bigl\{ \rho e^{-(\tau+\pi)\sigma^2/\rho}
\Phi({D\over2};i{\sigma^2\over2\rho},1+{\tau\over\pi}) \nn\\
&&+\sigma^2\pi e^{-(\tau+\pi)/\rho}
\Phi({D\over2}-1;{i\over2\rho},1+{\tau\over\pi}) \label{I1zeroT} \\
&&-\left.\sigma^2{2\pi^2\over\rho}e^{-(\tau+\pi)/\rho}
\Phi({D\over2}-2;{i\over2\rho},1+{\tau\over\pi}) -(\tau\ra-\tau)\Bigr\}
\right]\ , \nn \\
I_1(0)&=&\rho^{D\over2}\int_0^{\pi/2}d\tau\cot\tau\cdot\left[
\tau^{-D\over2} + \pi^{-D\over2}\Bigl\{
\zeta({D\over2},1+{\tau\over\pi}) 
-\zeta({D\over2},1-{\tau\over\pi})\Bigr \}\right] \ , 
\label{I1zeroT0}
\eeqa
and differentiating $I_1$, 
\beqa
I_2(\sigma)&=&\left.{\partial\over\partial\sigma^2}
   I_1(\sigma)\right|_{\sigma=0} \nn \\
&=& \rho^{D/2-1}\int_0^{\pi/2}d\tau\cot\tau\cdot\left[
\tau^{1-D/2}\Bigl\{-1+e^{-\tau/\rho}(1-{2\tau\over\rho})
\Bigr\}\right.\nn\\
&&+ \pi^{1-D/2}\Bigl\{ -\zeta\Bigl({D\over2}-1,1+{\tau\over\pi}\Bigr) 
  + e^{-(\tau+\pi)/\rho}
\Phi\Bigl({D\over2}-1;{i\over2\rho},1+{\tau\over\pi}\Bigr) \nn\\
&&-\left.{2\pi\over\rho}e^{-(\tau+\pi)/\rho}
\Phi\Bigl({D\over2}-2;{i\over2\rho},1+{\tau\over\pi}\Bigr) 
-(\tau\ra-\tau)\Bigr\} \right]\ . \label{i2above}
\eeqa 
We obtain the following critical $\rho$-values as the solutions of  \eq{criticalE} with 
\eq{i2above}
\beq
\rho_c=\left\{ 
\begin{array}{ll}
 0.583    &\quad\mbox{for} \quad D=2  \\
 1.238   &\quad \mbox{for} \quad D=3 \\ 
 3.394   &\quad \mbox{for} \quad D=3.5\ , \\ 
\end{array}\right.               \label{rhoc}
\eeq
and we verified that the real part of the potential given by \eq{RVR} 
with \eq{I1zeroT} and \eq{I1zeroT0} behaves consistently. Here we 
add a few remarks on these results. From $D=2$ to $D=3.3$, whose 
$\rho_c=1.965$, all $\rho_c$-values excellently coincide with the 
solutions of the following analytic equation
\beqa
0&=& 2 -(2\rho)^{{D\over2}-1}\zeta(2-{D\over2})
\cos\Bigl[{\pi\over4}(D-2)\Bigr]           \label{dualgap} \\
&&+\mbox{Re}\,\left[
(2i\rho)^{{D\over2}-1}\zeta\Bigl(2-{D\over2},1-{i\over2\rho}\Bigr)
+(2i\rho)^{{D\over2}-2}(D-4)\zeta\Bigl(3-{D\over2},1-{i\over2\rho}\Bigr)
\right] \ ,\nn
\eeqa
and this still exhibits a good 
agreement~\footnote{The deviation of $\rho_c$ grows as $D$ increasing 
in the region $D>3.5$. We believe this is related to a singular 
problem in numerical computations like encountered in sect. 3.2.}
at $D=3.5$ ($\rho_c=3.395$). 
We found this equation from \eq{gapm} through rotating $\xi\ra i\rho$, 
taking real part and the limit $m\ra0$. The divergence stemming from 
$m\ra0$ now drops off since it becomes a pure imaginary number by the 
rotation. Similarly, we verify that those critical values are 
consistent with the following potential derived by the 
same rotation procedure from \eq{magpote};
\beqa
V(\sigma;\rho)={\sigma^2\over2\lambda_R}
&+&\mbox{Re}\,\Biggl[\,{\trone i\rho \over2(4\pi)^{D/2}}\Bigl[\,
2\sigma^2(2i\rho)^{{D\over2}-2}\zeta(2-{D\over2},1-{i\over2\rho})
\Gamma(2-{D\over2})  \nn\\ 
&-& 4\sigma^2(2i\rho)^{{D\over2}-3}
\zeta(3-{D\over2},1-{i\over2\rho})\Gamma(3-{D\over2}) \label{ddpp} \\
&+& 2(2i\rho)^{{D\over2}-1}\Gamma(1-{D\over2})\Bigl\{\,
 \zeta(1-{D\over2},1-{i\sigma^2\over2\rho})
 -\zeta(1-{D\over2}) \, \Bigr\} \,\Bigr]\, \Biggr] \ .\nn
\eeqa
Note that this rotation procedure is not always allowed in general 
cases (for example in the thermodynamic part ${\tilde V}$), because 
we have used the expansion \eq{coth} (or the formula \eq{formula1} 
derived from \eq{coth}), which is not valid for a complex analytic 
function. In this sense, the above coincidences are non-trivial 
results (although they are rather trivial before the expansion 
is applied).

Now we discuss the phase structures in the $\rho$-$T$-$\mu$ space. 
In Fig.9, we depict phase diagrams for several $\mu$ values in the 
case of $D=2$. As mentioned above, the vacuum state is not invariant 
at $T=\mu=0$. In this sense, the critical value $\rho_c$ divides a 
phase between massive states (broken phase) and quasi-stational 
states (quasi-symmetric phase), and the total probability of decay 
per volume is given by \cite{KKM} 
\beq
\Gamma = -\lim_{\rho\ra\rho_c} 2\sigma{\rm Im}\,V'
(\sigma;\rho,\infty,0) = {\rho_c\over\pi}.
\eeq
The similar situation is expected for finite $\mu$ (with $T=0$) 
as well, and the $\Gamma$ changes into $\Gamma+\Delta\Gamma$, 
where $\Delta\Gamma$ is formally given by 
\beq
\Delta\Gamma=-\lim_{\rho\ra\rho_c}2\sigma\mbox{Im}\,
{\tilde V}'(\sigma;\rho,\infty,\mu)\ ,
\eeq
although we have not estimated its explicit value yet. 
On the other hand, as discussed previously, the chiral symmetry can 
be restored (i.e., $\Gamma=0$) when both of $T$ and $\mu$ are finite. 

In Figs. 10 and 11, the cases of $D=3$ and $4-\epsilon$ ($\epsilon=0.5$) 
are illustrated. As shown in \eq{impozero}, these cases have a stable 
(chiral symmetric) vacuum at $T=\mu=0$. Also, in the situation of 
finite $T$ and $\mu$, the chiral symmetry is restored similarly 
to the $D=2$ case. In each dimension, the critical surface goes 
down to a low temperature region as $\mu$ increasing. 
We could not find any tricritical point within a region of 
significant numerical data. In particular the computation did not 
work well below around $T\cong0.1$, and we hence extrapolate the 
critical lines in Figs.9-11.

\vspace{1cm}
\begin{minipage}[t]{7cm}  
\epsfxsize=8.5cm
\epsfbox{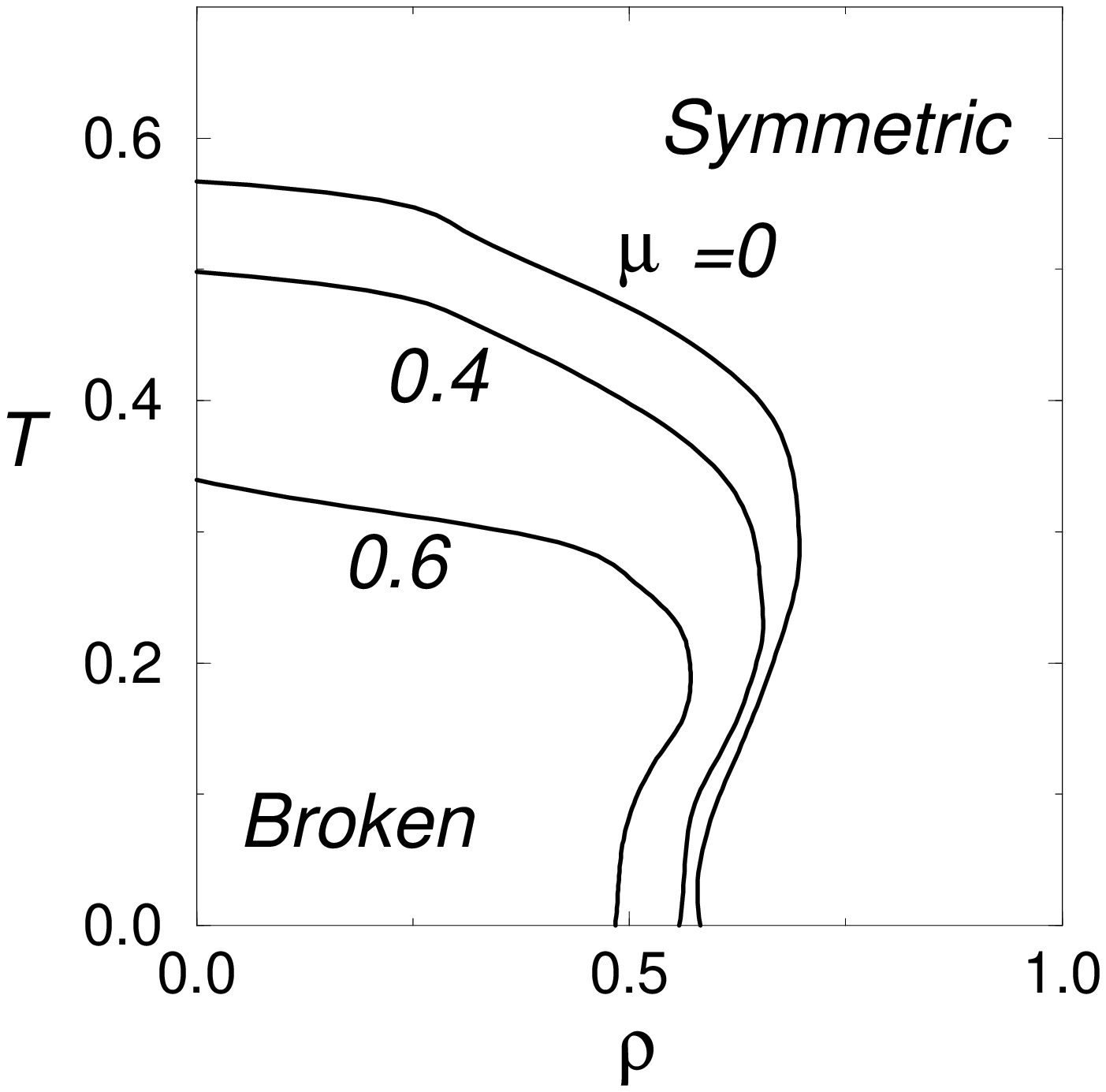} 
  {\bf Fig. 9:} The critical lines for electric dominant cases in $D=2$. 
\end{minipage}
\hspace{0cm}
\begin{minipage}[t]{7cm}
\epsfxsize=8.5cm
\epsfbox{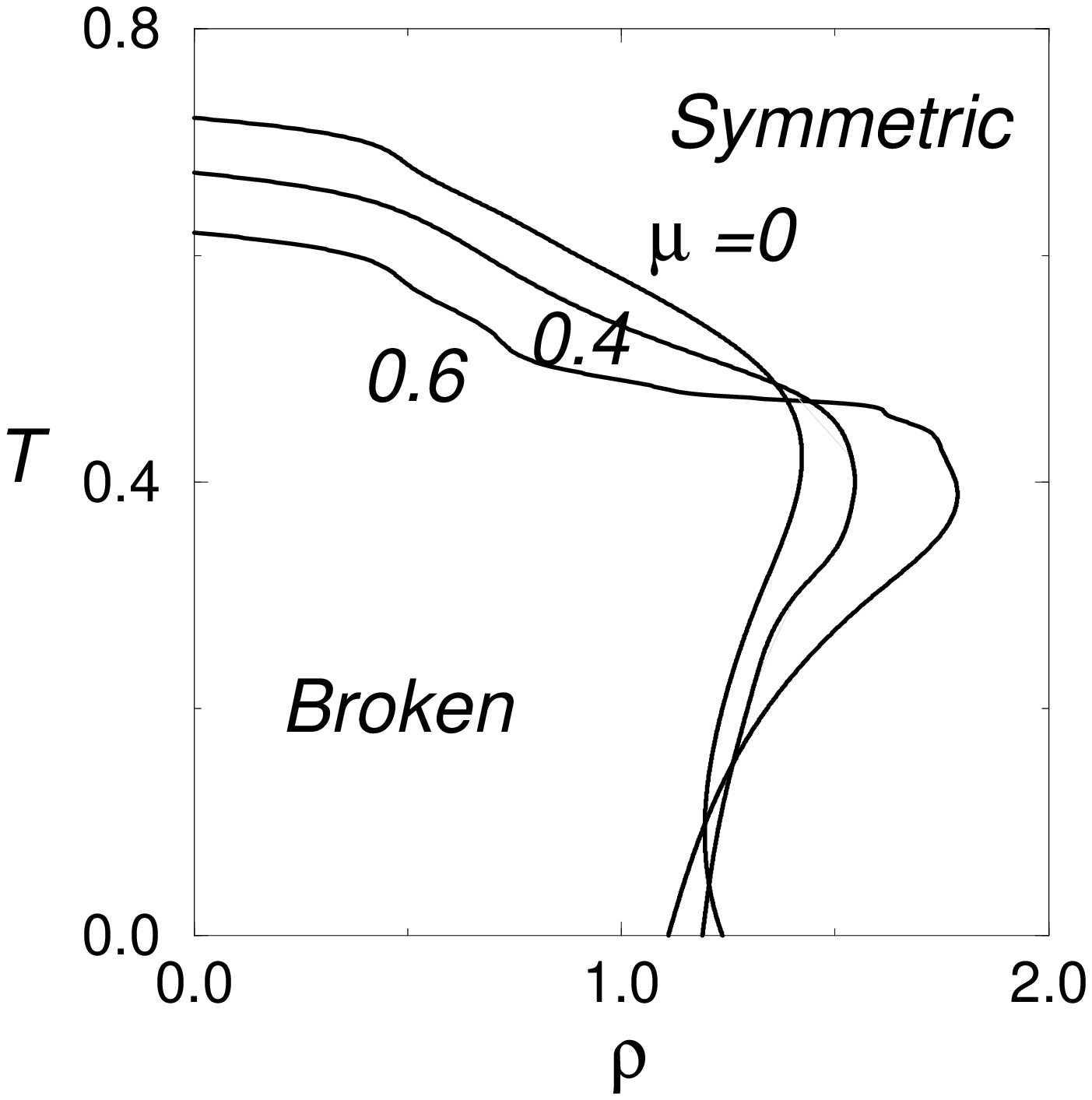}
  {\bf Fig. 10:} The critical lines for electric dominant cases in $D=3$.
\end{minipage}
\vspace{5mm}

\vspace{1cm}
\begin{minipage}[t]{7cm}  
\epsfxsize=8.5cm
\epsfbox{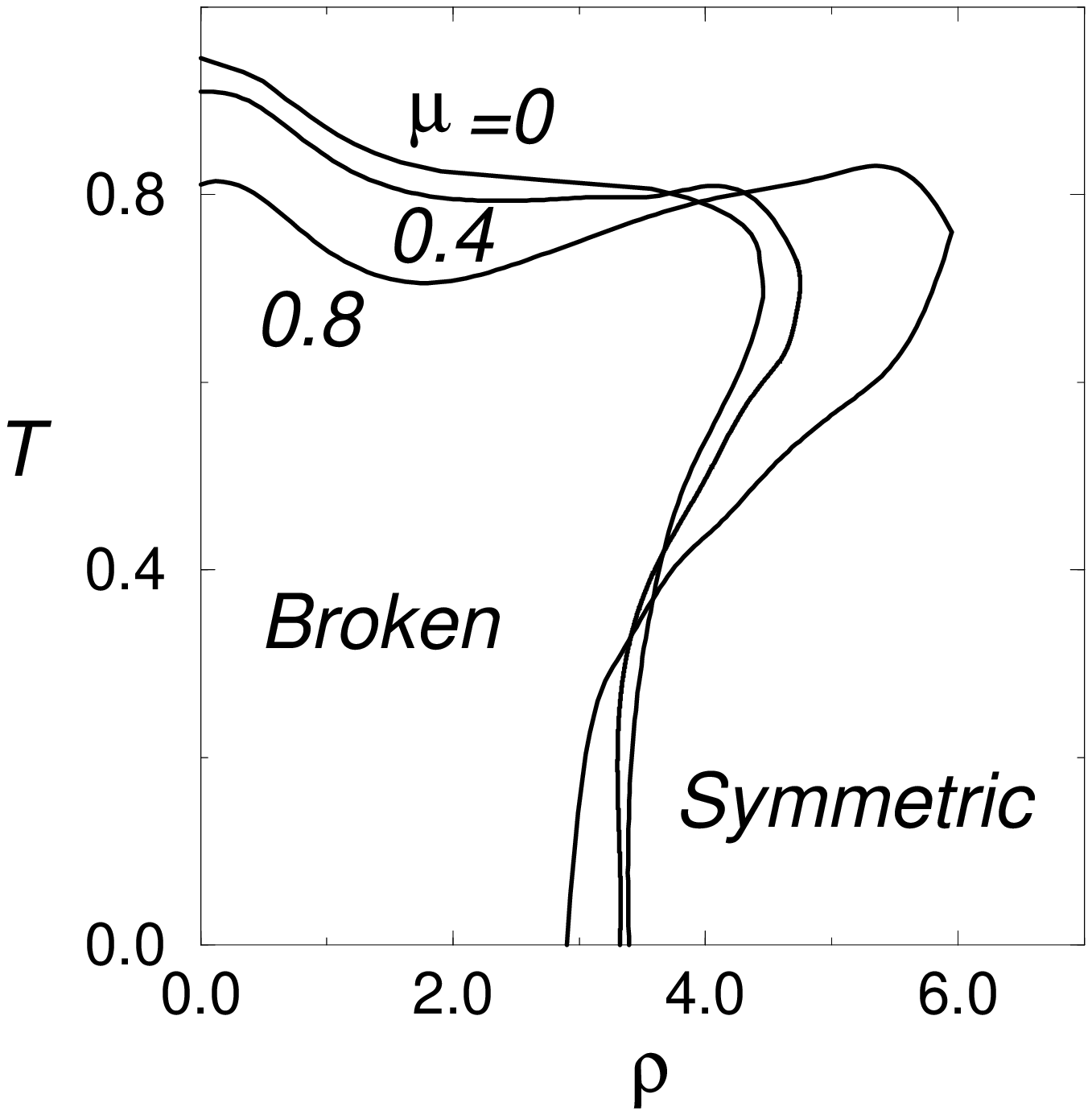}  
  {\bf Fig. 11:}   The critical lines for electric dominant cases in 
                   $D=4 - \epsilon$, where $\epsilon$
                  is chosen to be $0.5$.
\end{minipage}
\vspace{5mm}

\section{Conclusion and discussion}
\setcounter{equation}{0}
\indent

%
In this paper, we found various interesting properties of the 
thermodynamic effective potential of the large-$N$ Gross-Neveu model 
coupled to constant electromagnetic fields perpendicular 
to each other: worldline approach, magnetic discretization, 
electromagnetic duality, critical equations and 
chiral phase structures. 

First, we summarize the analytic properties of the effective 
potentials. In sect.2, we presented a new derivation of the effective 
potential from the point of the worldline approach developed in 
recent years \cite{Str,SS}. The similar formulation had been applied 
to constant curvature cases in \cite{KS}, but we could not clarify 
the correspondence to other calculations. However, we emphasize that 
it is interesting that our method proposed in sect.2 leads to the 
exact results \cite{GMST,Anders}. Another spectacular of our potential 
can be seen in sect.3. Our potential reveals the role of $\xi$ as a 
discrete momentum unit even in the situations under finite 
temperature with keeping $\mu$ finite, and in particular, the 
discrete potential of $D=3$ completely resembles to the continuum 
potential \cite{IKM}. In other words, our $D$-dimensional 
thermodynamic effective potential is composed of the sum over 
discrete energy spectra $\epsilon_n=\pm\sqrt{m^2+2n\xi}$, $n\in Z$. 
We also argued how to take the $\beta\ra\infty$ limit of the 
potential with $\xi$ and $\mu$ kept finite, and we described the 
precise continuum limit $\xi\ra0$, which gives rise to the known 
results of $\mu_c$. It means $\mu$ is properly taken into account 
in our derivation. 

We found in sect.4 the tri-criticality condition for the imaginary 
part of the electric dominant effective potential, and also found 
the analytic expressions \eq{dualgap} and \eq{ddpp} which are dual 
to the magnetic dominant effective potential and its gap equation 
at $T=\mu=0$. When the effective potential possesses the imaginary 
part, the vanishing conditions on the first and second derivatives 
of ${\rm Re}V$ are not sufficient for defining second order critical 
and tricritical points. It is necessary for second order critical 
points to satisfy the condition \eq{discri2}, which is a vanishing 
condition on the chiral mass term $\sigma{\rm Im}V'$. 
Similarly, \eq{discri3}, $\sigma^{-1}{\rm Im}V'=0$, should be 
satisfied at a tricritical point as well. These vanishing conditions 
on ${\rm Im}V'$ may be regarded as a discrete version of the real 
part conditions \eq{cond1} and \eq{cond2}.

Based on all these properties, we investigated the phase structures 
in various cases. First, in sect.3, we showed the phase diagrams 
for magnetic dominant cases, and observed the strong catalyst 
effects of magnetic fields. When $\beta=\infty$ ($\mu\not=0$ in 
$D=3$ and $3.5$), the symmetry restorations are not the second but 
the first order. In these cases, the chiral symmetry becomes broken 
in a strong magnetic field region. This property is the same for 
finite $\beta$ (with $\mu\not=0$) also. These features do not change 
if we formally extend to the $2< D <3$ regions, from 
which we actually got some insights toward a region $D$ being near 
to 4. As pointed out in \cite{DGN}, the effective potential of $D=2$ 
by itself resembles to an electrostatic potential problem of the charge 
distribution $s\coth s$ along a one-dimensional line. It means that 
such an electric distribution rather acts as a "magnetic" 
catalyst, if we could realize the distribution in a one-dimensional 
laboratory. 

The phase diagram analyses for the electric dominant cases are more 
complicated. First we have to test whether the imaginary part of the 
effective potential vanishes as $\sigma$ goes to zero (in other words, 
the chiral mass term $\sigma\langle{\bar\psi}\psi\rangle$ vanishes or 
not, q.v. \eq{kkmcri}). At $T=\mu=0$, this can be easily checked by 
integrating \eq{impozero} w.r.t. $\sigma$, and it shows that the 
${\rm Im}\,V(\sigma)$ behaves like $\sigma^{D-2}$. For $2< D<4$, the 
chiral symmetry gets restored. For $D=2$, the ${\rm Im}\,V'$ diverges 
as $\sigma\ra0$, and the imaginary part ${\rm Im}\,V$ never 
vanishes \cite{KKM}.  

However, at finite $\beta$ and $\mu$, we showed that the chiral 
symmetry is restored in $2\leq D<4$ based on the same criterion. 
This suggests that a symmetrization effect by high temperature can 
be stronger than the vacuum instability by an external field. As to 
$\rho$-$\mu$ plane, we have not yet succeeded in formulating the 
$T=0$ limit with $\mu$ being kept finite (In this sense, the 
achievement of the magnetic case was remarkable). When 
the fermion density is large, a density effect might enhance the 
chiral symmetry, and consequently, the instability would be suppressed. 
In order to address this question (at $T=0$), it is necessary to study 
the 1'st order critical equation like done in~\eq{ximu1cri}. 
An interesting problem is whether or not a dual relation like the one 
between \eq{gapm} and \eq{dualgap} holds in this case, because it 
contains a non-trivial part, the thermodynamic quantity ${\tilde V}$, 
which has non-analytic dependence on $\mu$. 

\hspace*{18pt}

\appendix

\section{A proof of (3.14)}
\setcounter{equation}{0}
\indent

First, we show how \eq{j} and \eq{m} reproduce \eq{c2line} and 
\eq{tricripoint} respectively. Then we present a proof of \eq{obeta0}.
In the limit $\xi\ra0$, \eq{j} becomes \eq{c2line} in the 
following way
\beqa
0&=&\integ ds s^{-D/2}\left[e^{-s}-e^{s\mu^2}{\sqrt{4\pi s}\over\beta}
\teta\right]         \label{defC2} \\
&=&\Gamma(1-{D\over2})-2{\sqrt{4\pi}\over\beta}\integ ds s^{1-D\over2}
\mbox{Re}\,\sum_{n=1}^\infty\exp\Bigl[
-s {4\pi^2\over\beta^2}(n-{1\over2}+i{\beta\mu\over2\pi})^2\Bigr]\nn\\
&=&\Gamma(1-{D\over2})-2{\sqrt{4\pi}\over\beta}
\left({2\pi\over\beta}\right)^{D-3}\Gamma({3-D\over2})\mbox{Re}\,
\sum_{n=0}^\infty(n+{1\over2}+i{\beta\mu\over2\pi})^{D-3}\nn\\
&=&\Gamma(1-{D\over2})-{2\over\sqrt{\pi}}
\left({2\pi\over\beta}\right)^{D-2}\Gamma({3-D\over2})\mbox{Re}\,
\zeta\Bigl(3-D,{1\over2}+i{\beta\mu\over2\pi}\Bigr)\ , \nn
\eeqa
and similarly \eq{m} reads 
\beqa
0&=&\integ ds s^{1-{D\over2}}e^{s\mu^2}{\sqrt{4\pi s}\over\beta}
\teta \nn\\
&=&{2\over\sqrt{\pi}}
\left({2\pi\over\beta}\right)^{D-4}\Gamma({5-D\over2})\mbox{Re}\,
\zeta\Bigl(5-D,{1\over2}+i{\beta\mu\over2\pi}\Bigr)\ . \nn
\eeqa

Let us turn to the proof of \eq{obeta0} making use of the above 
calculation. Recalling the definition of ${\cal O}_\beta(\sigma)$ 
by \eq{obeta}, and using \eq{transj}, we have
\beq
{\cal O}_\beta(\sigma)=\integ ds s^{-D/2}e^{-s\sigma^2}\left[
\Theta_4\Bigl(i{\beta\mu\over2\pi},
i{\beta^2\over4\pi s}\Bigr)-1\right]\ .
\eeq
The top line of \eq{defC2} can also be written in the form 
\beq
0=\integ ds s^{-D/2}\left[\,e^{-s}-
\Theta_4\Bigl(i{\beta\mu\over2\pi},i{\beta^2\over4\pi s}\Bigr)\,\right]\ . 
\label{proofkey}
\eeq
If $2<D<4$, the following integral is convergent
\beq
\integ ds s^{-D/2}(e^{-s}-1)= 
{1\over1-{D\over2}}\integ ds s^{1-{D\over2}}e^{-s}=
\Gamma(1-{D\over2})\ ,
\eeq
and therefore \eq{proofkey} amounts to
\beq
0=\Gamma(1-{D\over2})-{\cal O}_\beta(0)\ .
\eeq
Comparing the RHS of this equation with the bottom line of \eq{defC2}, 
we have proven \eq{obeta0}.
\section{The $\xi\ra0$ limit of (3.22)}
\setcounter{equation}{0}
\indent

We prove that \eq{zeropoteD} is reduced to \eq{zeropote} in the 
limit $\xi\ra0$. Obviously, the 1'st term 
${\cal O}_\infty(0)$ in \eq{zeropoteD} drops in $\xi\ra0$, since 
${\cal O}_\infty(0)$ possesses no dependence on $\xi$. Hence, we 
have only to prove the following equality 
\beq
\lim_{\xi\ra0}\sum_{l=1}^L\xi\int_0^{\mu-\sqrt{2l\xi}}
(t^2+2t\sqrt{2l\xi})^{D-3\over2}dt={\mu^D\over D(D-1)} \ . 
\label{lhsb1}
\eeq

Under the correspondence \eq{corres}, LHS of the above equation 
can be transformed into 
\beqa
\mbox{LHS}\ \eq{lhsb1}&=&
\int_0^\mu kdk\int_0^{\mu-k}(t^2+2kt)^{D-3\over2}dt 
=\int_0^\mu dt\int_0^{\mu-t}k(t^2+2kt)^{D-3\over2}dk \nn\\
&=& \int_0^\mu dtt^{D-1}\int_0^{(\mu-t)/t}k(1+2k)^{D-3\over2}dk\ ,
\eeqa
and applying the formula 3.194.1 \cite{GR}
\beq
\int_0^ux^{\mu-1}(1+\beta x)^{-\nu}dx ={1\over\mu}u^\mu 
F(\nu,\mu;1+\mu;-\beta u)\ ,\qquad |\mbox{arg}(1+\beta u)|<\pi,
\quad \mbox{Re}\,\mu>0 \ ,   \label{Grad31941}
\eeq
we get
\beq
\mbox{LHS}\ \eq{lhsb1} = 
{1\over2}\int_0^\mu t^{D-3}(\mu-t)^2F\Bigl({3-D\over2},2;3;
-2{\mu-t\over t}\Bigr)\ .
\eeq
Now changing the integration variable, $t=2\mu/(w+2)$,
\beq
\mbox{LHS}\ \eq{lhsb1} = 
\mu^D 2^{D-3}\integ dw w^2 (w+2)^{-1-D}
F\Bigl({3-D\over2},2;3;-w\Bigr)
\eeq
and applying the formula 7.512.10 \cite{GR}
\beq
\integ x^{\gamma-1}(x+z)^{-\sigma}F(\alpha,\beta;\gamma;-x)dx=
{\Gamma(\gamma)\Gamma(\alpha-\gamma+\sigma)\Gamma(\beta-\gamma+\sigma)
\over\Gamma(\sigma)\Gamma(\alpha+\beta-\gamma+\sigma)} \ ,
\eeq
(where $\mbox{Re}\,\gamma>0$, $\mbox{Re}\,\alpha-\gamma+\sigma>0$, 
$\mbox{Re}\,\beta-\gamma+\sigma>0$, $|\mbox{arg}\,z|<\pi$), we find 
\beq
\mbox{LHS}\ \eq{lhsb1}
= \mu^D2^{D-3}{\Gamma(3)\Gamma({D-1\over2})\Gamma(D)\over
\Gamma(D+1)\Gamma({D+3\over2})}F\Bigl({D-1\over2},D;{D+3\over2};-1\Bigr)\ .
\label{LHSfinal}
\eeq
Using a Kummer transformation to make the argument of hypergeometric 
function from $-1$ to $1/2$ and the following formula
\beq
F(\alpha,1-\alpha;\gamma;{1\over2})={2^{1-\gamma}\Gamma(\gamma)\sqrt{\pi}
\over\Gamma({\alpha+\gamma\over2})\Gamma({\gamma-\alpha+1\over2})}\ ,
\eeq
we have
\beq
F({D-1\over2},D;{3+D\over2};-1)={2^{1-D}\Gamma({3+D\over2})\over
\Gamma({D+1\over2})}\ .
\eeq
Finally substituting this into \eq{LHSfinal}, the proof ends
\beq
\mbox{LHS}\ \eq{lhsb1}
=\mu^D{1\over4}{\Gamma(3)\Gamma(D)\Gamma({D-1\over2})\over
\Gamma(D+1)\Gamma({D+1\over2})} = {\mu^D\over D(D-1)}\ .
\eeq

\section{The $\xi\ra0$ limit of (3.28)}
\setcounter{equation}{0}
\indent

We prove \eq{ximu1cri} goes to \eq{c1line} in the limit $\xi\ra0$. 
Eq.\eq{ximu1cri} is nothing but the sum of two quantities 
\beq
0=V(m;\xi)+{\tilde V}(m;\xi,\infty,\mu)\ , \label{vmxi}
\eeq
where $V(m,\xi)$ is given by \eq{vms}, and ${\tilde V}(m;\xi,\infty,\mu)$ 
is given by RHS of \eq{zeropoteD}. According to the result of Appendix B, 
we have
\beq
\lim_{\xi\ra0}{\tilde V}(m;\xi,\infty,\mu)=
{\trone2\sqrt{\pi}\over(4\pi)^{D/2}\Gamma({D-1\over2})}
{\mu^D\over D(D-1)}\ .           \label{resofapb}
\eeq

Let us consider the limit of another quantity $V(m;\xi)$. In the 
situation of $\xi\ra0$, three terms of zeta function on RHS of 
\eq{vms} only survive
\beqa
V(m;\xi)&\cong& {\trone\xi\over2(4\pi)^{D/2}}\left[
2(2\xi)^{{D\over2}-2}m^2\Gamma(2-{D\over2})\sum_{n=0}^\infty 
{1\over(n+1+m^2/2\xi)^{2-D/2}}\right.          \label{vsim}\\
&&\quad +\left.2(2\xi)^{{D\over2}-1}\Gamma(1-{D\over2})
\sum_{n=0}^\infty\Bigl\{{1\over(n+1+m^2/2\xi)^{1-D/2}}
-{1\over(n+1)^{1-D/2}}\Bigr\}\right]\ ,\nn
\eeqa
and we adopt the following continuum limit as $\xi\ra0$,
\beq
(n+1)2\xi \quad\ra\quad x\ ,\qquad\quad 
2\xi\sum_{n=0}^\infty \quad \ra\quad \integ dx\ .
\eeq
The limit of \eq{vsim} is therefore led to the following 
\beqa
\lim_{\xi\ra0}V(m;\xi)&=&{\trone\over2(4\pi)^{D/2}}\integ dx 
\left[{-m^2\Gamma(2-{D\over2})\over(x+m^2)^{2-D/2}}+
\Gamma(1-{D\over2})\Bigl\{(x+m^2)^{{D\over2}-1}
-x^{{D\over2}-1}\Bigr\}\right]\nn\\
&=&{\trone\over2(4\pi)^{D/2}}\Gamma(1-{D\over2})\left.
\Bigl[(x+m^2)^{D/2}({2\over D}-{m^2\over x+m^2})
-{2\over D}x^{D/2}\Bigr]\right|_0^\infty\nn\\
&=&{\trone\over2(4\pi)^{D/2}}\Gamma(1-{D\over2})(1-{2\over D})m^D\ .
\label{C5}
\eeqa
Combining \eq{resofapb} and \eq{C5}, the limit of \eq{vmxi} turns 
out to be 
\beq
{1\over2}\Gamma(1-{D\over2})(1-{2\over D})m^D +
{2\sqrt{\pi}\over\Gamma({D-1\over2})}{\mu^D\over D(D-1)}=0\ .
\eeq
This is equivalent to \eq{c1line}, and the proof ends.

\section{The explicit forms of $I_2(\sigma)$ and $I_3(\sigma)$}
\setcounter{equation}{0}
\indent

We write down explicit expressions for $I_2(\sigma)$ and $I_3(\sigma)$, 
which are defined by
\beqa
I_2(\sigma)&=&\left.{\partial\over\partial\sigma^2}
             I_1(\sigma)\right|_{\sigma=0}\ ,\\
I_3(\sigma)&=& \left.{\partial^2\over\partial(\sigma^2)^2}
             I_1(\sigma)\right|_{\sigma=0}\ .
\eeqa
$I_2(\sigma)$ is composed of the sum of the following two quantities:
\beqa
\left.{\partial\over\partial\sigma^2}A(\sigma)
\right|_{\sigma=0} &=&
2\sum_{k=0}^{\infty}\zeta(2k)
\left({\rho\over\pi}\right)^{2k}
\left[2{\sqrt{4\pi}\over\beta}\mbox{Re}\,\sum_{m=0}^\infty
(\pi{2m+1\over\beta}+i\mu)^{D-3-4k}\right.\nn\\
&&\quad \times \gamma\Bigl({3-D\over2}+2k,
{\pi\over2\rho}(\pi{2m+1\over\beta}+i\mu)^2\Bigr) \\
&&\quad -\left.\gamma(1-{D\over2}+2k,{\pi\over2\rho})
+2\gamma(2-{D\over2}+2k,{\pi\over2\rho}) \nn\right]\ ,
\eeqa
and 
\beqa
\left.{\partial\over\partial\sigma^2}B(\sigma)
\right|_{\sigma=0} \hs{-1}&=&\hs{-1}
\int_0^{\pi/2}d\tau\cot\tau\cdot\left[
-2{\sqrt{4\pi}\over\beta}\left({\rho\over\pi}\right)^{D-3\over2}
\mbox{Re}\sum_{m=0}^\infty \exp\Bigl[-{\tau+\pi\over\rho} 
(\pi{2m+1\over\beta}+i\mu)^2\Bigr]\right.  \nn\\
&&\quad\times\Phi\Bigl({D-3\over2};{i\over2\rho}
(\pi{2m+1\over\beta}+i\mu)^2,1+{\tau\over\pi}\Bigr) \nn\\
&&\quad+\left({\rho\over\pi}\right)^{{D\over2}-1}
e^{-(\tau+\pi)/\rho}
\Phi({D\over2}-1;{i\over2\rho},1+{\tau\over\pi}) \nn\\
&&\quad-\left.2\left({\rho\over\pi}\right)^{{D\over2}-2}
e^{-(\tau+\pi)/\rho}
\Phi({D\over2}-2;{i\over2\rho},1+{\tau\over\pi}) 
-(\tau\ra -\tau) \right] \ .
\eeqa
And similarly, $I_3(\sigma)$ is composed of:
\beqa
\left.{\partial^2\over\partial(\sigma^2)^2}A(\sigma)
\right|_{\sigma=0}\hs{-1} &=&\hs{-1}
-\,4{\sqrt{4\pi}\over\beta}\sum_{k=0}^{\infty}\zeta(2k)
\left({\rho\over\pi}\right)^{2k}
\mbox{Re}\,\sum_{m=0}^\infty
(\pi{2m+1\over\beta}+i\mu)^{D-5-4k} \nn\\
&&\quad \times \gamma\Bigl({5-D\over2}+2k,
{\pi\over2\rho}(\pi{2m+1\over\beta}+i\mu)^2\Bigr) \ ,
\eeqa
and 
\beqa
\left.{\partial^2\over\partial(\sigma^2)^2}B(\sigma)
\right|_{\sigma=0} \hs{-1.5}&=&\hs{-1.5}
2{\sqrt{4\pi}\over\beta} \left({\rho\over\pi}\right)^{D-5\over2}
\int_0^{\pi/2}d\tau\cot\tau\cdot
\mbox{Re}\left[\sum_{m=0}^\infty \exp\Bigl[-{\tau+\pi\over\rho} 
(\pi{2m+1\over\beta}+i\mu)^2\Bigr]\right.  \nn\\
&&\quad\times\left.\Phi\Bigl({D-5\over2};{i\over2\rho}
(\pi{2m+1\over\beta}+i\mu)^2,1+{\tau\over\pi}\Bigr) 
-(\tau\ra-\tau)\,\right] \ .\nn\\
\eeqa

Our notation for the Lerch transcendental $\Phi(z;s,a)$ is 
\beq
\Phi(z;s,a)=\sum_{n=0}^\infty{e^{2\pi ins}\over(n+a)^z}\ ,
\qquad \mbox{Re}\,z>1\ ,  \label{lerch}
\eeq
and a Lerch transformation is then expressed by 
\beqa
(2\pi)^z\Phi(1-z;s,a)&=&\Gamma(z)\left[
\exp\Bigl[\pi i({z\over2}-2as)\Bigr]\Phi(z;-a,s)\right.\nn\\
&&\ +\left.\exp\Bigl[\pi i(-{z\over2}+2a(1-s))\Bigr]
\Phi(z;a,1-s)\right]\ .
\eeqa
Taking $a=1$, we have
\beq
e^{2\pi is}\Phi(1-z;s,1)=(2\pi)^{-z}\Gamma(z)\Bigl[
e^{{\pi i\over2}z}\zeta(z,s) +
e^{-{\pi i\over2}z}\zeta(z,1-s)\, \Bigr] \ . \label{Ltrans}
\eeq
%

\end{document}